\documentclass[]{iopart}

\usepackage{iopams}
\usepackage{bm,cite}
\usepackage[dvipdfm]{graphicx}

\def\be{\begin{equation}}
\def\ee{\end{equation}}
\def\ba{\begin{eqnarray}}
\def\ea{\end{eqnarray}}

\newtheorem{definition}{Definition}[section]

\newtheorem{prop}[definition]{Proposition}

\newcommand{\hook}
{\raisebox{-0.35ex}{\makebox[0.6em][r]{\scriptsize $-$}}
\hspace{-0.15em}\raisebox{0.25ex}{\makebox[0.4em][l]{\tiny $|$}}}

\begin{document}

\begin{flushright}
KOBE-TH-14-08\\
OCU-PHYS 411\\
\end{flushright}

\title{A simple test for spacetime symmetry}

\author{Tsuyoshi Houri$^{1,2}$ and Yukinori Yasui$^3$}

\address{%
$^1$Department of Physics, Kobe University, Kobe, 657-8501 Japan,\\
$^2$Osaka City University Advanced Mathematical Institute (OCAMI), Osaka, 558-8585 Japan,\\
$^3$Department of Mathematics and Physics, Osaka City University, Osaka, 558-8585 Japan
}%

\begin{abstract}
This paper presents a simple method for investigating spacetime symmetry for a given metric.
The method makes use of the curvature conditions that are obtained from the Killing equations.
We use the solutions of the curvature conditions to compute an upper bound on the number of Killing vector fields,
as well as Killing-Yano tensors and closed conformal Killing-Yano tensors.
We also use them in the integration of the Killing equations.
By means of the method, we thoroughly investigate Killing-Yano symmetry of type D vacuum solutions
such as the Kerr metric in four dimensions.
The method is also applied to a large variety of physical metrics in four and five dimensions.
\end{abstract}


\maketitle

\section{Introduction}
Spacetime symmetry has played an important role in general relativity,
particularly isometry that is described by Killing vector fields
\be
 \nabla_\mu \xi_\nu + \nabla_\nu \xi_\mu = 0 \,. \label{KV}
\ee
The analysis of the equation leads to the well-known result
that if the spacetime dimension is given by $n$,
the metric admits at most $n(n+1)/2$ linearly independent Killing vector fields.
Only Minkowski, de Sitter and anti-de Sitter spacetimes are maximally symmetric,
which admit the maximum number of Killing vector fields.
At the same time, it is not always easy to find Killing vector fields for a given metric
because one needs to solve coupled partial differential equations obtained from (\ref{KV}).
Alternatively, a Killing vector field $\partial/\partial x$ can be identified
if all metric components are independent of a coordinate $x$.
However, finding such a coordinate is still difficult.

Meanwhile, Killing-Yano (KY) tensors have been recognised to be describing hidden symmetry of spacetimes
because thanks to such symmetry many complicated physical problems become tractable.
For instance, the Kerr spacetime admits a nondegenerate rank-2 KY tensor
that guarantees integrability of the Hamilton-Jacobi equation for geodesics
\cite{Carter:1968,Floyd:1973,Penrose:1973,Hughston:1973b}.
KY tensors were originally introduced in \cite{Yano:1952}
as a generalisation of Killing vector fields to higher-rank antisymmetric tensors
\be
 \nabla_\mu k_{\nu_1\nu_2\dots \nu_p} + \nabla_{\nu_1}k_{\mu\nu_2\dots \nu_p} = 0 \,. \label{KY}
\ee
In four dimensions, if a spacetime admits a nondegenerate rank-2 KY tensor,
the Hamilton-Jacobi equation for geodesics,
the Klein-Gordon and the Dirac equations can be solved by separation of variables \cite{Carter:1977,Benenti:1979,Carter:1979}.
The similar properties can be seen also in higher dimensions
(see reviews \cite{Frolov:2008,Yasui:2011} and references therein).
However, it is as difficult to find KY tensors as Killing vector fields.

The purpose of this paper is to present a simple method for investigating KY tensors for a given metric.
Since rank-1 KY tensors are 1-forms dual to Killing vector fields,
the method can be applied to Killing vector fields, too.
We also deal with the Hodge duals of KY tensors,
which are known as closed conformal Killing-Yano (CCKY) tensors.
The idea is based on the work of U.\ Semmelmann \cite{Semmelmann:2002}.
It was shown that one can introduce a connection
on the vector bundle $E^p(M)\equiv \Lambda^pT^*M \oplus \Lambda^{p+1}T^*M$, known as a Killing connection,
whose parallel sections are one-to-one corresponding to rank-$p$ KY tensors.
In this paper, using the Killing connection, we calculate the curvature on the vector bundle $E^p(M)$.
From the curvature and its covariant derivative, we obtain some curvature conditions
that provide necessary conditions for the parallel sections.
Solving the curvature conditions,
the number of linearly independent solutions puts an upper bound on the number of KY tensors.

A feature of the method is that the curvature conditions are obtained as algebraic equations,
which enables us to compute the upper bound for any metric.
In Sec.~3, we actually compute the upper bound on the number of Killing vector fields on the Kerr spacetime.
We will see that the Kerr spacetime admits exactly two Killing vector fields
without solving the Killing equation.
Another feature is that the solution of the curvature conditions
gives an ansatz for solving the original differential equations (\ref{KV}) and (\ref{KY}).
As we will see in Sec.~3, the Killing equation for the Kerr metric becomes tractable with such an ansatz.

This paper is organised as follows:
In Sec.~2, we begin with the familiar discussion on the maximum number of Killing vector fields.
Following \cite{Semmelmann:2002}, we introduce the Killing connection and calculate its curvature.
Hence, we obtain the curvature conditions that provide an upper bound on the number of Killing vector fields.
Sec.~3 shows how to exploit the obtained curvature conditions.
As an example, we actually investigate Killing vector fields on the Kerr spacetime.
In Sec.~4, we extend the discussion in Sec.~2 to higher-rank KY tensors.
Similar to Killing vector fields, we introduce a Killing connection.
Calculating its curvature, we derive curvature conditions that provide an upper bound on the number of KY tensors.
We also discuss the curvature conditions on CCKY tensors.
In Sec.~5, only using type D vacuum conditions,
we thoroughly investigate KY symmetry of type D vacuum solutions in four dimensions.
Sec.~6 also investigates KY tensors for some physical metrics in four and five dimensions,
where we only list the results (see Table 1, 2 and 3).
Sec.~7 is devoted to discussion and conclusions.

\section{Curvature conditions on Killing vector fields}
As is well-known, the maximum number of linearly independent Killing vector fields
on an $n$-dimensional spacetime $M$ with a Lorentzian metric $g_{\mu\nu}$ is obtained
by the following discussion (e.g., see the Wald's book \cite{Wald:1984}).
The Killing equation (\ref{KV}) is written as
\be
 \nabla_\mu \xi_\nu = L_{\mu\nu} \,, \label{KV1}
\ee
where $L_{\mu\nu}=\nabla_{[\mu}\xi_{\nu]}$
and $\nabla$ is the Levi-Civita connection.
Taking the covariant derivative, eq.\ (\ref{KV1}) leads to
\be
 \nabla_\mu L_{\nu\rho} = - R_{\nu\rho\mu}{}^\sigma\xi_\sigma \,, \label{KV2}
\ee
where $R_{\mu\nu\rho}{}^\sigma$ is the Riemann curvature.
Eqs.\ (\ref{KV1}) and (\ref{KV2}) show that a Killing vector field $\xi^\mu$ is determined
by the initial values of $\xi_\mu$ and $L_{\mu\nu}$ at a point on $M$.
Since $L_{\mu\nu}$ is antisymmetric, one can provide at most $n(n+1)/2$ data at each point,
which give the maximum number of Killing vector fields.

We consider the covariant derivative of eq.\ (\ref{KV2}).
Then, we obtain the curvature condition
\ba
 & (\nabla_\mu R_{\rho\sigma\nu}{}^\lambda)\xi_\lambda - (\nabla_\nu R_{\rho\sigma\mu}{}^\lambda)\xi_\lambda \nonumber\\
 & - R_{\mu\nu\rho}{}^\lambda L_{\sigma\lambda} + R_{\mu\nu\sigma}{}^\lambda L_{\rho\lambda} 
   - R_{\rho\sigma\mu}{}^\lambda L_{\nu\lambda} + R_{\rho\sigma\nu}{}^\lambda L_{\mu\lambda} = 0 \,, \label{CurvCond2}
\ea
where we have used eq.\ (\ref{KV1}).
Since the condition is no longer differential equations
but linear algebraic equations for $\xi_\mu$ and $L_{\mu\nu}$,
it provides restrictions on the values of $\xi_\mu$ and $L_{\mu\nu}$ at each point on $M$.
In four dimensions, there are 20 equations for 10 functions.
In $n$ dimensions, there are $n^2(n^2-1)/12$ equations for $n(n+1)/2$ functions.

Furthermore, taking the covariant derivative of eq.\ (\ref{CurvCond2}), we obtain the curvature condition
\be \eqalign{
 \nabla_\kappa \nabla_{[\mu}R_{|\rho\sigma|\nu]}{}^\lambda \xi_\lambda
 + R_{\mu\nu[\rho}{}^\delta R_{\sigma]\delta\kappa}{}^\lambda \xi_\lambda
 + R_{\rho\sigma[\mu}{}^\delta R_{\nu]\delta\kappa}{}^\lambda \xi_\lambda  \\
 + \nabla_{[\mu}R_{|\rho\sigma|\nu]}{}^\lambda L_{\kappa\lambda}
 - \nabla_\kappa R_{\mu\nu[\rho}{}^\lambda L_{\sigma]\lambda}
 - \nabla_\kappa R_{\rho\sigma[\mu}{}^\lambda L_{\nu]\lambda} =0 \,, } \label{CurvCond4}
\ee
where the derivative terms $\nabla_\mu\xi_\nu$ and $\nabla_\mu L_{\nu\rho}$ have been replaced by $\xi_\mu$ and $L_{\mu\nu}$
with the help of eqs.\ (\ref{KV1}) and (\ref{KV2}).
It provides further restrictions on the values of $\xi_\mu$ and $L_{\mu\nu}$ at each point on $M$.

One can take further derivatives of the curvature condition and,
in principle, one could take the infinite number of the derivatives.
However, in this paper, we only deal with the curvature condition (\ref{CurvCond2})
and its first derivative (\ref{CurvCond4}).
Those conditions are strong enough to restrict the values of $\xi_\mu$ and $L_{\mu\nu}$
at least for metrics that we investigate in this paper, as we will see later.
If the curvature conditions only have the trivial solution, $\xi_\mu=L_{\mu\nu}=0$,
one can conclude that the metric admits no Killing vector field.

\section{Killing vector fields on the Kerr spacetime}
We investigate Killing vector fields on the Kerr spacetime
by use of the curvature conditions (\ref{CurvCond2}) and (\ref{CurvCond4}).
For the Kerr metric, we begin with the metric form \cite{Carter:1968},
\be \eqalign{
 ds^2 =& \frac{r^2+p^2}{{\cal Q}}dr^2+\frac{r^2+p^2}{{\cal P}}dp^2 \nonumber\\
       & -\frac{{\cal Q}}{r^2+p^2}(d\tau-p^2d\sigma)^2
          +\frac{{\cal P}}{r^2+p^2}(d\tau+r^2d\sigma)^2 \,, } \label{Kerr_metric}
\ee
where
\be
 {\cal Q} = r^2 - 2mr + a^2 \,, \quad {\cal P} = a^2 - p^2 \,. \label{Kerr_function}
\ee
For later calculation, we introduce an orthonormal frame
\be
\eqalign{
 {\bm e}^1 = \frac{dr}{\sqrt{Q_1}} \,, \quad {\bm e}^2 = \frac{dp}{\sqrt{Q_2}} \,, \\
 {\bm e}^{\hat{1}} = \sqrt{Q_1}(d\tau-p^2d\sigma) \,, \quad
 {\bm e}^{\hat{2}} = \sqrt{Q_2}(d\tau+r^2d\sigma) \,,
} \label{OrthonoramalFrame}
\ee
where
\be
 Q_1 = \frac{{\cal Q}}{r^2+p^2} \,, \quad Q_2 = \frac{{\cal P}}{r^2+p^2} \,.
\ee
In such a frame, the metric is written as ${\bm g} = {\bm e}^1{\bm e}^1+{\bm e}^2{\bm e}^2
-{\bm e}^{\hat{1}}{\bm e}^{\hat{1}}+{\bm e}^{\hat{2}}{\bm e}^{\hat{2}}$.
Since all the metric components are independent of the coordinates $\tau$ and $\sigma$,
$\partial/\partial\tau$ and $\partial/\partial\sigma$ are Killing vector fields.
The dual 1-forms are given by
\ba
 {\bm \xi}_1 = \sqrt{Q_1}{\bm e}^{\hat{1}}+\sqrt{Q_2} {\bm e}^{\hat{2}} \,, \label{Killing_One_Form_1} \\
 {\bm \xi}_2 = p^2\sqrt{Q_1}{\bm e}^{\hat{1}}-r^2\sqrt{Q_2} {\bm e}^{\hat{2}} \,, \label{Killing_One_Form_2}
\ea
which are known Killing 1-forms on the Kerr spacetime.

For the metric, we first solve the curvature condition (\ref{CurvCond2}) for $\xi_\mu$ and $L_{\mu\nu}$.
Suppose that ${\bm \xi} = a_1 \,{\bm e}^1 + a_2 \,{\bm e}^2 + a_3 \,{\bm e}^{\hat{1}} + a_4 \,{\bm e}^{\hat{2}}$ and
${\bm L} = a_5 \,{\bm e}^1\wedge {\bm e}^2 + a_6 \,{\bm e}^1\wedge {\bm e}^{\hat{1}} + a_7 \,{\bm e}^1\wedge {\bm e}^{\hat{2}}
+ a_8 \,{\bm e}^2\wedge {\bm e}^{\hat{1}} + a_9 \,{\bm e}^2\wedge {\bm e}^{\hat{2}} + a_{10} \,{\bm e}^{\hat{1}}\wedge {\bm e}^{\hat{2}}$
where $a_i$ $(i=1,\dots,10)$ are unknown functions.
Solving eq.\ (\ref{CurvCond2}), we obtain $a_1=a_2=a_5=a_{10}=0$ and
\ba
 a_7 = \frac{2 r}{r^2+p^2} \left( a_4 \sqrt{Q_1}-a_3 \sqrt{Q_2} \right) \,, \\
 a_8 = \frac{2 p}{r^2+p^2} \left( a_4 \sqrt{Q_1}-a_3 \sqrt{Q_2} \right) \,,
\ea
hence, $\xi_\mu$ and $L_{\mu\nu}$ take the form
\ba
 {\bm \xi} =& a_3 \,{\bm e}^{\hat{1}} + a_4 \,{\bm e}^{\hat{2}} \,, \label{sol1} \\
 {\bm L} =& a_6 \,{\bm e}^1\wedge {\bm e}^{\hat{1}} + \frac{2 r}{r^2+p^2} \left( a_4 \sqrt{Q_1}
            - a_3 \sqrt{Q_2} \right) \,{\bm e}^1\wedge {\bm e}^{\hat{2}} \nonumber\\
          & + \frac{2 p}{r^2+p^2} \left( a_4 \sqrt{Q_1}-a_3 \sqrt{Q_2} \right) \,{\bm e}^2\wedge {\bm e}^{\hat{1}} 
            + a_9 \,{\bm e}^2\wedge {\bm e}^{\hat{2}} \,, \label{sol2}
\ea
where $a_3$, $a_4$, $a_6$ and $a_9$ are arbitrary functions.
Since the solution of the curvature condition (\ref{CurvCond2}) is parametrised
by four parameters at each point,
the Kerr metric admits at most four Killing vector fields.

Furthermore, in addition to the curvature condition (\ref{CurvCond2}),
we solve the curvature condition (\ref{CurvCond4}).
Then, we find the solution
\ba
 a_6 = 2 \left(\partial_r \sqrt{Q_1}\right) a_3 - \frac{2r\sqrt{Q_2}}{r^2+p^2} a_4 \,, \\
 a_9 = - \frac{2p\sqrt{Q_1}}{r^2+p^2} a_3 + 2\left(\partial_p \sqrt{Q_2}\right) a_4 \,.
\ea
Since the independent parameters have reduced to two parameters $a_3$ and $a_4$,
the Kerr metric admits at most two Killing vector fields.
As we already have two Killing vector fields,
we can conclude that the Kerr metric admits exactly two Killing vector fields.

Finally, we attempt to solve the Killing equation (\ref{KV}).
We already know of course that the two Killing 1-forms are given by (\ref{Killing_One_Form_1}) and (\ref{Killing_One_Form_2}).
However, even if we did not know the Killing 1-forms, we can obtain them as follows.
Since the Killing 1-forms on the Kerr spacetime must take the form (\ref{sol1}),
we make use of the form as an ansatz for solving the Killing equation.
Then, it comes to be easy to solve the Killing equation.
The solution is given by
\be
 a_3 = (c_1+c_2 p^2) \sqrt{Q_1} \,, \quad a_4 = (c_1-c_2 r^2) \sqrt{Q_2} \,,
\ee
hence, we have
\be \eqalign{
 {\bm \xi} =& c_1 \left( \sqrt{Q_1}{\bm e}^{\hat{1}} + \sqrt{Q_2} {\bm e}^{\hat{2}} \right) \\
            & + c_2 \left( p^2 \sqrt{Q_1}{\bm e}^{\hat{1}} - r^2 \sqrt{Q_2} {\bm e}^{\hat{2}} \right) \,, }
\ee
where $c_1$ and $c_2$ are constants.
This is indeed the linear combination of the known Killing 1-forms
(\ref{Killing_One_Form_1}) and (\ref{Killing_One_Form_2}).

\section{Generalisation}
The discussion in Sec.~2 can be interpreted as the following.
From eqs.\ (\ref{KV1}) and (\ref{KV2}), one can introduce a connection ${\cal D}_\mu$
on the vector bundle $E^1(M)\equiv T^*M \oplus \Lambda^2T^*M$,
\be
 {\cal D}_\mu
 \left(
\begin{array}{c}
 \xi_\nu \\
 L_{\nu\rho}
\end{array}
\right)
 \equiv \left(
\begin{array}{c}
 \nabla_\mu\xi_\nu-L_{\mu\nu} \\
 \nabla_\mu L_{\nu\rho} + R_{\nu\rho\mu}{}^\sigma{}\xi_\sigma
\end{array}
\right) \,, \label{KC}
\ee
where $\xi_\mu$ is a 1-form and $L_{\mu\nu}$ is a 2-form.
The connection is known as a Killing connection.
It is manifest that if a section $\hat{\xi}_A$ of $E^1(M)$ is given by a Killing vector field $\xi^\mu$
and its exterior derivative $\nabla_{[\mu}\xi_{\nu]}$, then $\hat{\xi}_A = (\xi_\mu,\nabla_{[\mu}\xi_{\nu]})$ satisfies
\be
 {\cal D}_\mu \hat{\xi}_A = 0 \,, \label{parallel_eq}
\ee
which means that $\hat{\xi}_A$ is a parallel section of $E^1(M)$.
Conversely, one can demonstrate that
if $\hat{\xi}_A=(\xi_\mu,L_{\mu\nu})$ is a parallel section of $E^1(M)$,
$\xi^\mu$ is a Killing vector field and $L_{\mu\nu}$ is its exterior derivative, $L_{\mu\nu}=\nabla_{[\mu}\xi_{\nu]}$.
It follows that Killing vector fields on $M$ are in one-to-one correspondence with
parallel sections of $E^1(M)$.
Since the number of linearly independent parallel sections is bound by the rank of $E^1(M)$,
the number of linearly independent Killing vector fields is also bound by the rank of $E^1(M)$,
which is given by $n(n+1)/2$.

Calculating the curvature of the Killing connection ${\cal D}_\mu$, 
we obtain some conditions for the parallel sections $\hat{\xi}_A$ of $E^1(M)$.
Since we have eq.\ (\ref{parallel_eq}), the parallel sections $\hat{\xi}_A$ satisfy the curvature condition
\be
  {\cal R}_{\mu\nu A}{}^B\hat{\xi}_B 
  \equiv ({\cal D}_\mu{\cal D}_\nu-{\cal D}_\nu{\cal D}_\mu )\hat{\xi}_A = 0 \,, \label{CurvCond}
\ee
where ${\cal R}_{\mu\nu A}{}^B$ is called a Killing curvature.
The Killing curvature gives linear maps $\hat{\xi}_A\to\hat{\xi}^\prime_A={\cal R}_{\mu\nu A}{}^B\hat{\xi}_B$
for any choice of $\mu$ and $\nu$ at each point on $M$.
Moreover, taking the covariant derivative of the curvature condition (\ref{CurvCond}) on $E^1(M)$,
we obtain further condition
\be
 ({\cal D}_\mu {\cal R}_{\nu\rho A}{}^B) \hat{\xi}_B = 0 \,,
\ee
where ${\cal D}_\mu {\cal R}_{\nu\rho A}{}^B$ is defined by (\ref{CovDer_Curvature}).
Since those conditions are algebraic equations for $\hat{\xi}_A$,
they give restrictions on the values of $\hat{\xi}_A$ at each point on $M$.
Hence, investigating the linearly independent solutions of the curvature conditions,
we obtain an upper bound on the number of linearly independent Killing vector fields.

\subsection{Curvature conditions on Killing-Yano tensors}
The discussion about Killing vector fields can be naturally generalised
to higher-rank KY tensors and CCKY tensors.
For the purpose, we slightly change our notation.
Let $(M,{\bm g})$ be an $n$-dimensional Riemannian or Lorentzian manifold
and $\nabla$ be the Levi-Civita connection.
We work in a local orthonormal frame of $TM$ denoted by $\{{\bm X}_a\}$
and its dual frame of $T^*M$ denoted by $\{{\bm e}^a\}$.
Namely, they satisfy ${\bm X}_a\hook {\bm e}^b=\delta_a^b$ where $\hook$ is the interior product.
The Latin indices $a,b,\dots$ range from $1$ to $n$.
To deal with Riemannian and Lorentzian metrics simultaneously,
we define the matrix $\eta_{ab}={\bm g}({\bm X}_a,{\bm X}_b)$ which is diagonal with entries $\pm1$.
The signature is $(+,+,\dots,+)$ for Riemannian or $(-,+,\dots,+)$ for Lorentzian metrics.
We also define ${\bm X}^a=\eta^{ab}{\bm X}_b$ and ${\bm e}_a=\eta_{ab}{\bm e}^b$,
where $\eta^{ab}$ is the inverse of $\eta_{ab}$.
For a vector field ${\bm V}=V^a{\bm X}_a$, we introduce the dual 1-form ${\bm V}^*=V^a{\bm e}_a$.
In other words, ${\bm V}^* = {\bm g}({\bm V}, - )$.

A rank-$p$ KY tensor (or a KY $p$-form) ${\bm k}$ is defined as a $p$-form satisfying
\be
 \nabla_{\bm X}{\bm k} = \frac{1}{p+1}{\bm X}\hook d{\bm k} \,, \label{KYeq}
\ee
for any vector field ${\bm X}$.
Covariantly differentiating (\ref{KYeq}), we obtain 
\be
 \nabla_{\bm X}(d{\bm k}) = \frac{p+1}{p}R^+({\bm X}){\bm k} \,, \label{KYeq2}
\ee
where
\be
 R^+({\bm X}) \equiv {\bm e}^a\wedge R({\bm X},{\bm X}_a) \,.
\ee
The Riemann curvature is defined by
\be
 R({\bm X},{\bm Y}) \equiv \nabla_{\bm X}\nabla_{\bm Y} - \nabla_{\bm Y}\nabla_{\bm X} 
                          - \nabla_{[{\bm X},{\bm Y}]} \,. \label{CurvatureTensor}
\ee
In accordance with eqs.\ (\ref{KYeq}) and (\ref{KYeq2}),
one can introduce a connection ${\cal D}$
on the vector bundle $E^p(M)\equiv \Lambda^pT^*M\oplus \Lambda^{p+1}T^*M$ \cite{Semmelmann:2002},
\be
 {\cal D}_{\bm X}
\left(
\begin{array}{c}
{\bm \omega} \\
{\bm \eta}
\end{array}
\right)
\equiv \nabla_{\bm X} \left(
\begin{array}{c}
{\bm \omega} \\
{\bm \eta}
\end{array}
\right)- \Gamma({\bm X})\left(
\begin{array}{c}
{\bm \omega} \\
{\bm \eta}
\end{array}
\right) \,, \label{KillingConnectionKY}
\ee
where ${\bm \omega}$ is a section of $\Lambda^pT^*M$,
${\bm \eta}$ is a section of $\Lambda^{p+1}T^*M$ and
\be
 \Gamma({\bm X}) \equiv
\left(
\begin{array}{cc}
 0 &\displaystyle{\frac{1}{p+1}}{\bm X}\hook \\
 \displaystyle{\frac{p+1}{p}}R^+({\bm X}) &0
\end{array}
\right) \,.
\ee
If a section $\hat{{\bm \omega}}=({\bm \omega},{\bm \eta})$ of $E^p(M)$ is given by a KY $p$-form ${\bm \omega}={\bm k}$
and its exterior derivative ${\bm \eta}=d{\bm k}$, then it satisfies the parallel equation
\be
 {\cal D}_{\bm X} \hat{{\bm \omega}} = 0 \,. \label{paralleleqKY}
\ee
Conversely, if $\hat{{\bm \omega}}=({\bm \omega},{\bm \eta})$ is a parallel section of $E^p(M)$
then ${\bm \omega}$ is a KY $p$-form and ${\bm \eta}$ is its exterior derivative, ${\bm \eta}=d{\bm \omega}$.
It follows that KY $p$-forms on $M$ are in one-to-one correspondence
with parallel sections of $E^p(M)$.
Hence, the maximum number of KY p-forms is bound by the rank of $E^p(M)$ \cite{Semmelmann:2002},
 which is given by
\be
 \textrm{rank} ~E^p(M) =
\left(
\begin{array}{c}
n  \\
p 
\end{array}
\right)+
\left(
\begin{array}{c}
n  \\
p+1 
\end{array}
\right)=
\left(
\begin{array}{c}
n+1  \\
p+1 
\end{array}
\right) \,. \label{MB_KY}
\ee
The equality is attained if a spacetime is maximally symmetric.
Note that when we take $p=1$,
eqs.\ (\ref{KYeq}) and (\ref{KYeq2}) are equivalent to eqs.\ (\ref{KV1}) and (\ref{KV2}).
Eqs.\ (\ref{KillingConnectionKY}) and (\ref{paralleleqKY}) correspond to eqs.\ (\ref{KC}) and (\ref{parallel_eq}), respectively.
The maximum number (\ref{MB_KY}) becomes $n(n+1)/2$ for $p=1$.

As before, we calculate the curvature of the Killing connection (\ref{KillingConnectionKY}) by
\be
 {\cal R}({\bm X},{\bm Y}) \equiv {\cal D}_{\bm X}{\cal D}_{\bm Y}-{\cal D}_{\bm Y}{\cal D}_{\bm X}
                                 - {\cal D}_{[{\bm X},{\bm Y}]} \,,
\ee
on the vector bundle $E^p(M)$.
A straightforward calculation leads to the Killing curvature written by
\be
 {\cal R}({\bm X},{\bm Y})
 = \left(
\begin{array}{cc}
 N_{11}({\bm X},{\bm Y}) &0 \\
 N_{21}({\bm X},{\bm Y}) &N_{22}({\bm X},{\bm Y})
\end{array}
\right) \,. \label{KCurv_KY1}
\ee
The entries are given by
\be \eqalign{
 N_{11}({\bm X},{\bm Y}) = R({\bm X},{\bm Y}) + \frac{1}{p}\Big\{{\bm X}\hook R^+({\bm Y})-{\bm Y}\hook R^+({\bm X})\Big\} \,, \\
 N_{21}({\bm X},{\bm Y}) = -\frac{p+1}{p}\Big\{(\nabla_{\bm X} R)^+({\bm Y})-(\nabla_{\bm Y}R)^+({\bm X})\Big\} \,, \\
 N_{22}({\bm X},{\bm Y}) = R({\bm X},{\bm Y}) + \frac{1}{p}\Big\{R^+({\bm X})({\bm Y}\hook \bullet)-R^+({\bm Y})({\bm X}\hook \bullet)\Big\} \,,
} \label{KillingCurvatureKY}
\ee
where
\be
 (\nabla_{\bm X}R)^+({\bm Y}) = {\bm e}^a\wedge (\nabla_{\bm X} R)({\bm Y},{\bm X}_a) \,.
\ee
Since we have eq.\ (\ref{paralleleqKY}),
the parallel sections $\hat{{\bm \omega}}=({\bm \omega},{\bm \eta})$ of $E^p(M)$ satisfy the curvature condition
\be
 {\cal R}({\bm X},{\bm Y}) \hat{{\bm \omega}} = 0 \,, \label{CC_KY}
\ee
which is equivalent to the conditions
\ba
 N_{11}({\bm X},{\bm Y}) {\bm \omega} = 0 \,, \label{CC_KY_1} \\
 N_{21}({\bm X},{\bm Y}) {\bm \omega} + N_{22}({\bm X},{\bm Y}) {\bm \eta} = 0 \,. \label{CC_KY_2}
\ea

To obtain further conditions for the parallel sections,
we calculate the covariant derivative of the curvature condition (\ref{CC_KY}).
Then, we obtain for parallel sections the condition
\be
 ({\cal D}_{{\bm X}}{\cal R})({\bm Y},{\bm Z}) \hat{{\bm \omega}} = 0 \,, \label{curv_cond_first}
\ee
where
\be \fl \qquad \eqalign{
 ({\cal D}_{{\bm X}}{\cal R})({\bm Y},{\bm Z})\hat{{\bm \omega}}
 &\equiv {\cal D}_{{\bm X}}({\cal R}({\bm Y},{\bm Z}) \hat{{\bm \omega}})
          - {\cal R}({\bm Y},{\bm Z})({\cal D}_X\hat{{\bm \omega}}) \\
 & ~~~~~ -{\cal R}(\nabla_{{\bm X}}{\bm Y},{\bm Z})\hat{{\bm \omega}}
   -{\cal R}({\bm Y},\nabla_{{\bm X}}{\bm Z})\hat{{\bm \omega}} \\
 &= (\nabla_{{\bm X}}{\cal R})({\bm Y},{\bm Z})\hat{{\bm \omega}}
    -\Gamma({\bm X}){\cal R}({\bm Y},{\bm Z})\hat{{\bm \omega}}
    +{\cal R}({\bm Y},{\bm Z})\Gamma({\bm X})\hat{{\bm \omega}} \,. } \label{CovDer_Curvature}
\ee
Note that the Bianchi identity is given by
\be
 ({\cal D}_{{\bm X}}{\cal R})({\bm Y},{\bm Z}) + ({\cal D}_{{\bm Y}}{\cal R})({\bm Z},{\bm X}) 
 + ({\cal D}_{{\bm Z}}{\cal R})({\bm X},{\bm Y}) = 0 \,.
\ee
More explicitly, eq.\ (\ref{curv_cond_first}) is given by
\ba
 (\nabla_{{\bm X}}N_{11})({\bm Y},{\bm Z}) {\bm \omega} 
  + \frac{1}{p+1}N_{11}({\bm Y},{\bm Z}) ({\bm X}\hook{\bm \eta}) = 0 \,, \label{CC_KY_3} \\
 (\nabla_{{\bm X}}N_{21})({\bm Y},{\bm Z}){\bm \omega}
  + \frac{p+1}{p}N_{22}({\bm Y},{\bm Z}) R^+({\bm X}) {\bm \omega}  \nonumber\\
  + (\nabla_{{\bm X}}N_{22})({\bm Y},{\bm Z}){\bm \eta}
  + \frac{1}{p+1}N_{21}({\bm Y},{\bm Z}) ({\bm X}\hook {\bm \eta}) = 0 \,, \label{CC_KY_4}
\ea
where we have used eqs.\ (\ref{paralleleqKY}), (\ref{CC_KY_1}) and (\ref{CC_KY_2}).

Note that when we take $p=1$, $N_{11}({\bm X},{\bm Y})=0$ holds identically for all ${\bm X}$ and ${\bm Y}$.
It follows consequently that eqs.\ (\ref{CC_KY_1}) and (\ref{CC_KY_3}) are automatically satisfied.
When we take $p=n-1$, $N_{22}({\bm X},{\bm Y})=0$ identically holds for all ${\bm X}$ and ${\bm Y}$.

\subsection{Curvature conditions on closed conformal Killing-Yano tensors}
Similarly, for a CCKY p-form ${\bm h}$, we have
\ba
 \nabla_{\bm X}{\bm h} = -\frac{1}{n-p+1}{\bm X}^* \wedge\delta {\bm h} \,, \label{CCKYeq}\\
 \nabla_{\bm X}(\delta {\bm h}) = -\frac{n-p+1}{n-p}R^-({\bm X}){\bm h} \,, \label{CCKYeq2}
\ea
where
\be
 R^-({\bm X}) \equiv {\bm X}^a\hook R({\bm X},{\bm X}_a) \,.
\ee
Hence, this time, one can introduce a connection ${\cal D}$
on the vector bundle $\tilde{E}^p(M)\equiv \Lambda^pT^*M\oplus \Lambda^{p-1}T^*M$,
\be
 {\cal D}_{\bm X}\left(
\begin{array}{c}
{\bm h} \\
\delta {\bm h}
\end{array}
\right)
\equiv \nabla_{\bm X}\left(
\begin{array}{c}
{\bm h} \\
\delta {\bm h}
\end{array}
\right) -\Gamma({\bm X}) \left(
\begin{array}{c}
{\bm h} \\
\delta {\bm h}
\end{array}
\right)
 \,,
\ee
where
\be
\Gamma({\bm X}) \equiv \left(
\begin{array}{cc}
 0 &\displaystyle{-\frac{1}{n-p+1}}{\bm X}^*\wedge \\
 \displaystyle{-\frac{n-p+1}{n-p}}R^-({\bm X}) &0
\end{array}
\right) \,. \label{KillingConnectionCCKY}
\ee
Similar to KY tensors, we can demonstrate that 
CCKY tensors on $M$ are in one-to-one correspondence with parallel sections of $\tilde{E}^p(M)$.
Hence, the number of CCKY p-forms is bound by the rank of $\tilde{E}^p(M)$,
 which is given by
\be
 \textrm{rank} ~\tilde{E}^p(M) =
\left(
\begin{array}{c}
n  \\
p 
\end{array}
\right)+
\left(
\begin{array}{c}
n  \\
p-1 
\end{array}
\right)=
\left(
\begin{array}{c}
n+1  \\
p 
\end{array}
\right) \,. \label{MB_CCKY}
\ee
Note that the number of rank-p CCKY tensors is same as that of rank-(n-p) KY tensors
because CCKY tensors are given as the Hodge duals of KY tensors.

Calculating the curvature of the Killing connection (\ref{KillingConnectionCCKY}),
we obtain the Killing curvature
\be
 {\cal R}({\bm X},{\bm Y})
 = \left(
\begin{array}{cc}
 M_{11}({\bm X},{\bm Y}) &0 \\
 M_{21}({\bm X},{\bm Y}) &M_{22}({\bm X},{\bm Y})
\end{array}
\right)
\ee
with the entries
\be \fl \qquad \eqalign{
 M_{11}({\bm X},{\bm Y}) = R({\bm X},{\bm Y}) + \frac{1}{n-p}\Big\{{\bm X}^*\wedge R^-({\bm Y})-{\bm Y}^*\wedge R^-({\bm X})\Big\} \,, \\
 M_{21}({\bm X},{\bm Y}) = \frac{n-p+1}{n-p}\Big\{(\nabla_{\bm X} R)^-({\bm Y})-(\nabla_{\bm Y}R)^-({\bm X})\Big\} \,, \\
 M_{22}({\bm X},{\bm Y}) = R({\bm X},{\bm Y}) + \frac{1}{n-p}\Big\{R^-({\bm X})({\bm Y}^*\wedge \bullet)-R^-({\bm Y})({\bm X}^*\wedge\bullet)\Big\} \,.
} \label{KillingCurvatureCCKY}
\ee
where
\be
 (\nabla_{\bm X}R)^-({\bm Y}) = {\bm e}^a\hook (\nabla_{\bm X} R)({\bm Y},{\bm e}_a) \,.
\ee
Hence, we obtain the curvature conditions
\ba
 M_{11}({\bm X},{\bm Y}) {\bm \omega} = 0 \,, \label{CC_CCKY_1} \\
 M_{21}({\bm X},{\bm Y}) {\bm \omega} + M_{22}({\bm X},{\bm Y}) {\bm \eta} = 0 \,. \label{CC_CCKY_2}
\ea
Furthermore, as before,
the covariant derivatives of the curvature conditions lead to
further conditions. Now, they are given by
\ba
 (\nabla_{{\bm X}}M_{11})({\bm Y},{\bm Z}) {\bm \omega} 
  - \frac{1}{n-p+1}M_{11}({\bm Y},{\bm Z}) ({\bm X}\hook{\bm \eta}) = 0 \,, \label{CC_CCKY_3} \\
 (\nabla_{{\bm X}}M_{21})({\bm Y},{\bm Z}){\bm \omega}
  - \frac{n-p+1}{n-p}M_{22}({\bm Y},{\bm Z}) R^-({\bm X}) {\bm \omega}  \nonumber\\
  + (\nabla_{{\bm X}}M_{22})({\bm Y},{\bm Z}){\bm \eta}
  - \frac{1}{n-p+1}M_{21}({\bm Y},{\bm Z}) ({\bm X}\hook {\bm \eta}) = 0 \,. \label{CC_CCKY_4}
\ea

\section{Killing-Yano tensors on type D vacuum spacetimes}
By use of the method that was shown in previous sections,
we shall reconsider KY symmetry of type D vacuum solutions.
The results in this section cover the previous works
\cite{Walker:1970,Hughston:1972,Hughston:1973,Collinson:1976,Stephani:1978,Hall:1987}
(see propositions \ref{prop1} to \ref{prop4}).
This section also aims to illustrate how simply the method enables us
to investigate KY symmetry of type D vacuum solutions.

We work in the Newman-Penrose formalism,
which introduces the complex null tetrad $\{{\bm X}_a\}=\{{\bm k},{\bm l},{\bm m},\bar{{\bm m}}\}$
that satisfies
\begin{equation}
 {\bm g}({\bm k},{\bm l}) = 1 \,, \quad {\bm g}({\bm m},\bar{{\bm m}}) = -1 \,. \label{NP}
\end{equation}
The basis ${\bm k}$ and ${\bm l}$ are real vector fields,
whereas ${\bm m}$ and $\bar{{\bm m}}$ are complex.
The complex conjugate of ${\bm m}$ is denoted by $\bar{{\bm m}}$.
The 1-forms $\{{\bm e}^a\}$, which satisfy ${\bm X}_a\hook {\bm e}^b=\delta_a^b$,
are given by $\{{\bm k}_*,{\bm l}_*,{\bm m}_*,\bar{{\bm m}}_*\}$.
Using the matrix $\eta_{ab}={\bm g}({\bm X}_a,{\bm X}_b)$,
we define 1-forms $\{{\bm e}_a\}=\{{\bm k}^*,{\bm l}^*,{\bm m}^*,\bar{{\bm m}}^*\}$ by ${\bm e}_a=\eta_{ab}{\bm e}^b$.
For the null tetrad, the spin coefficients are defined as usual:
\be \fl \quad \eqalign{
 \kappa = {\bm g}({\bm m},\nabla_{{\bm k}}{\bm k}) \,, \quad
 \sigma = {\bm g}({\bm m},\nabla_{{\bm m}}{\bm k}) \,, \quad
 \lambda = {\bm g}({\bm l}, \nabla_{\bar{{\bm m}}}\bar{{\bm m}}) \,, \quad
 \nu = {\bm g}({\bm l}, \nabla_{{\bm l}}\bar{{\bm m}}) \,, \\
 \rho = {\bm g}({\bm m}, \nabla_{\bar{{\bm m}}}{\bm k}) \,, \quad
 \mu = {\bm g}({\bm l}, \nabla_{{\bm m}}\bar{{\bm m}}) \,, \quad
 \tau = {\bm g}({\bm m}, \nabla_{{\bm l}}{\bm k}) \,, \quad
 \pi = {\bm g}({\bm l}, \nabla_{{\bm k}}\bar{{\bm m}}) \,, \\
 \alpha = \frac{1}{2}\Big\{{\bm g}({\bm l}, \nabla_{\bar{{\bm m}}}{\bm k}) + {\bm g}({\bm m}, \nabla_{\bar{{\bm m}}}\bar{{\bm m}})\Big\} \,, \quad
 \beta = \frac{1}{2}\Big\{{\bm g}({\bm l}, \nabla_{{\bm m}}{\bm k}) + {\bm g}({\bm m}, \nabla_{{\bm m}}\bar{{\bm m}})\Big\} \,, \\
 \epsilon = \frac{1}{2}\Big\{{\bm g}({\bm l}, \nabla_{{\bm k}}{\bm k}) + {\bm g}({\bm m}, \nabla_{{\bm k}}\bar{{\bm m}})\Big\} \,, \quad
 \gamma = \frac{1}{2}\Big\{{\bm g}({\bm l}, \nabla_{{\bm l}}{\bm k}) + {\bm g}({\bm m}, \nabla_{{\bm l}}\bar{{\bm m}})\Big\} \,, }
 \label{spincoefficients}
\ee
where the all spin coefficients are complex in general.

Firstly, we consider type D conditions.
There are two principal null directions of multiplicity two in type D spacetimes.
When the ${\bm k}$ and ${\bm l}$ are chosen along the two principal null directions,
type D conditions are given by a complex scalar function $\Psi\neq 0$,
\be \eqalign{ 
 C({\bm k},{\bm l},{\bm k},{\bm l})=-\Psi-\bar{\Psi} \,, \quad
 C({\bm k},{\bm l},{\bm m},\bar{{\bm m}})=\Psi-\bar{\Psi} \,, \\
 C({\bm k},{\bm m},{\bm l},\bar{{\bm m}})=\Psi \,, \quad
 C({\bm k},\bar{{\bm m}},{\bm l},{\bm m})=\bar{\Psi} \,, \\
 C({\bm m},\bar{{\bm m}},{\bm m},\bar{{\bm m}})=-\Psi-\bar{\Psi} \,, } \label{curvNP}
\ee
where $C$ is the Weyl curvature. The other components are vanishing.
In addition, we consider vacuum condition,
under which the Riemann curvature is equal to the Weyl curvature.
From the Goldberg-Sachs theorem, type D vacuum conditions lead to
\be
 \kappa = \sigma = \nu = \lambda = 0 \,. \label{Goldberg-Sachs}
\ee
Furthermore, making the boost and the rotation of the basis,
we can set $\epsilon=0$ while preserving (\ref{Goldberg-Sachs}).

All type D vacuum solutions in four dimensions
were obtained by W.\ Kinnersley \cite{Kinnersley:1969}.
He classified the solutions into four cases (case I--IV).
However, we do not use the explicit expressions of the solutions.
Only using type D vacuum conditions on the Weyl curvature and its covariant derivatives,
we compute the Killing curvature (\ref{KCurv_KY1})
on the vector bundle $E^p(M)$ (see appendix A for details).
Solving the curvature conditions (\ref{CC_KY}) and (\ref{curv_cond_first}),
the following propositions \ref{prop1} to \ref{prop4} are obtained.

\begin{prop} \label{prop1}
Every type D vacuum solution, with the exception of case III
of Kinnersley's classification \cite{Kinnersley:1969},
admits exactly one rank-2 Killing-Yano tensor.
Case III solution does not admit any rank-2 Killing-Yano tensor.
\end{prop}
\paragraph{Proof.}
The Killing curvature on $E^2(M)$ are given by (\ref{KC2_1})-(\ref{KC2_18}).
Solving the curvature conditions (\ref{CC2_0}),
the solution $\hat{\bm \omega}=({\bm \omega},{\bm \eta})$ is given at each point by
\ba
 {\bm \omega} &=& \omega_1 \,{\bm k}_* \wedge {\bm l}_*
                  + \omega_2 \,{\bm m}_* \wedge \bar{{\bm m}}_* \,, \\
 {\bm \eta} &=& \eta_1 \,{\bm k}_*\wedge {\bm l}_*\wedge {\bm m}_* + \eta_2 \,{\bm k}_*\wedge {\bm l}_*\wedge \bar{{\bm m}}_* \nonumber\\
            & & + \eta_3 \,{\bm k}_*\wedge {\bm m}_*\wedge \bar{{\bm m}}_* + \eta_4 \,{\bm l}_*\wedge {\bm m}_*\wedge \bar{{\bm m}}_* \,,
\ea
where $\omega_1$ and $\omega_2$ are free complex parameters and
\be \eqalign{
 \eta_1=-3 \tau(\omega_1-\omega_2) \,, \quad
 \eta_2=3 \pi(\omega_1-\omega_2) \,, \\
 \eta_3=3 \rho(\omega_1-\omega_2) \,, \quad
 \eta_4=-3 \mu(\omega_1-\omega_2) \,. } \label{D_KY5}
\ee
Imposing the reality conditions $\bar{{\bm \omega}}={\bm \omega}$ and $\bar{{\bm \eta}}={\bm \eta}$,
the parameters must satisfy the conditions
$\omega_1=\bar{\omega}_1$, $\omega_2=-\bar{\omega}_2$,
$\eta_2=\bar{\eta}_1$, $\eta_3=-\bar{\eta}_3$ and $\eta_4=-\bar{\eta}_4$.
The reality conditions together with (\ref{D_KY5}) lead to the conditions
\ba
 (\rho+\bar{\rho})\,\omega_1-(\rho-\bar{\rho})\,\omega_2=0 \,, \label{D_KY1}\\
 (\mu+\bar{\mu})\omega_1-(\mu-\bar{\mu})\omega_2=0 \,, \label{D_KY2}\\
 (\tau+\bar{\pi})\,\omega_1-(\tau-\bar{\pi})\,\omega_2=0 \,, \label{D_KY3}
\ea
where $\omega_1$ is a real and $\omega_2$ is a pure imaginary parameter.
For cases I-III ($\rho\neq0$),
the spin coefficients $\rho$, $\mu$, $\tau$ and $\pi$ satisfy the relations
(e.g., see \cite{Ishikawa:1982})
\ba
 \mu = -U \rho \,, \quad \rho \tau+\bar{\rho}\pi = 0 \,,
\ea
where $U$ is a certain real function.
Then, eq.\ (\ref{D_KY2}) is equivalent to (\ref{D_KY1}).
Eq.\ (\ref{D_KY3}) is written as 
\be
 (\rho\tau-\bar{\rho}\bar{\tau})\omega_1 
 - (\rho\tau+\bar{\rho}\bar{\tau})\omega_2 = 0 \,. \label{D_KY4}
\ee
For cases I and II, $\tau$ is pure imaginary
and hence eq.\ (\ref{D_KY4}) reduces to eq.\ (\ref{D_KY1}).
This implies that the case I and II solutions admit at most one rank-2 KY tensor.
On the other hand, we have two independent conditions (\ref{D_KY1}) and (\ref{D_KY4}) for case III,
to which the nonzero solution for $\omega_1$ and $\omega_2$ does not exist.
Thus, the case III solution does not admit any rank-2 KY tensor.
For case IV, we have $\rho=\mu=0$ \cite{Kinnersley:1969}.
Eqs.\ (\ref{D_KY1}) and (\ref{D_KY2}) become identities.
Eq.\ (\ref{D_KY3}) remains as the only equation to solve, so that
case IV solution admits at most one rank-2 KY tensor.
Since it is known that every type D vacuum solution, except for the case III solution,
admits (at least) a rank-2 KY tensor \cite{Walker:1970,Hughston:1972,Hughston:1973,Collinson:1976,Stephani:1978},
we arrive at the statement of the present proposition. \hfill$\Box$\\

\noindent{\bf Remark.}\quad
Proposition \ref{prop1} is consistent with the result of \cite{Hall:1987}.
In \cite{Hall:1987}, spacetimes admitting at least two rank-2 KY tensors were discussed in four dimensions.
Since type D vacuum solutions admit just one rank-2 KY tensor,
they are outside the latter calss of metrics.
When considering the Euclidean counterparts of the solutions,
we can construct self-dual Ricci-flat (hyper-K\"ahler) metrics for particular choice of the parameters.
In those cases, we obtain additional rank-2 KY tensors that are hyper-K\"ahler forms (see Sec.~6.3 for details).\\

\begin{prop} \label{prop2}
Case II and III solutions admit exactly two Killing vector fields,
whereas case I and IV admit exactly four Killing vector fields.
\end{prop}
\paragraph{Proof.}
The Killing curvature on $E^1(M)$ is given by (\ref{KC1_1})-(\ref{KC1_12}).
Solving the curvature condition (\ref{CC1_0}) obtained from the Killing curvature,
the parallel sections $\hat{\bm \omega}=({\bm \omega},{\bm \eta})$ of $E^1(M)$
are necessarily written at each point on $M$ as
\ba
 {\bm \omega} &=& \omega_1 \,{\bm k}_* + \omega_2 \,{\bm l}_*
                  + \omega_3 \,{\bm m}_* + \omega_4 \,\bar{{\bm m}}_*, \\
 {\bm \eta} &=& \eta_1 \,{\bm k}_*\wedge {\bm l}_* + \eta_2 \,{\bm k}_*\wedge {\bm m}_* 
                + \eta_3 \,{\bm k}_*\wedge \bar{{\bm m}}_* \nonumber\\
            & & + \eta_4 \,{\bm l}_*\wedge {\bm m}_* + \eta_5 \,{\bm l}_*\wedge \bar{{\bm m}}_* 
                + \eta_6 \,{\bm m}_*\wedge \bar{{\bm m}}_* \,,
\ea
where $\omega_i$ $(i=1,\cdots,4)$ and $\eta_j$ $(j=1,\cdots,6)$ are free complex parameters
with the constraints
\be \eqalign{
 \eta_2 = 2 \tau \omega_1-2 \rho \omega_3 \,, \quad
 \eta_5 = -2 \pi \omega_2+2 \mu \omega_4 \,, \\
 \mu \omega_1 -\rho \omega_2 - \pi \omega_3 + \tau \omega_4 = 0 \,. } \label{const_KY}
\ee
The reality conditions $\bar{{\bm \omega}}={\bm \omega}$ and $\bar{{\bm \eta}}={\bm \eta}$
imply that $\omega_1=\bar{\omega}_1$, $\omega_2=\bar{\omega}_2$, $\omega_4=\bar{\omega}_3$,
$\eta_1=\bar{\eta}_1$, $\eta_3=\bar{\eta}_2$, $\eta_4=\bar{\eta}_5$ and $\eta_6=-\bar{\eta}_6$.
Together with (\ref{const_KY}), the remaining degrees of freedom are given by four real parameters
for case II and IIIB, or given by five real parameters for case I, IIIA and IV.
In addition, solving the condition (\ref{CC_KY_4}) [cf.\ (\ref{CurvCond4})]
that is obtained from the covariant derivative
of the curvature condition (\ref{CC1_0}), we obtain the result of the proposition. \hfill$\Box$\\

\begin{prop} \label{prop3}
All type D vacuum solutions admit no rank-3 Killing-Yano tensor.
\end{prop}
\paragraph{Proof.}
The Killing curvature on $E^3(M)$ are given by (\ref{KC3_1})-(\ref{KC3_6}).
In particular, $N_{11}({\bm k},{\bm l})$ and $N_{11}({\bm m},\bar{{\bm m}})$
are given by diagonal matrices with nonzero entries that are independent of any spin coefficient.
It follows that their kernels are vanishing.
Hence, we find that there exists no rank-3 KY tensor. \hfill$\Box$\\

\noindent{\bf Remark.}\quad
Since the Hodge duals of KY tensors are CCKY tensors, 
propositions \ref{prop1} to \ref{prop3} hold
even if we replace the rank-$p$ KY tensors by rank-($4-p$) CCKY tensors.\\

A covariantly constant $p$-form is a rank-$p$ KY tensor and also a rank-$p$ CCKY tensor.
From proposition \ref{prop3}, we immediately find that all type D vacuum solutions do not admit
any covariantly constant 1-forms and 3-forms.
As was seen in the proof of proposition \ref{prop1},
KY 2-forms on type D vacuum solutions must take the form
\be
 {\bm \omega} = \omega_1 \,{\bm k}_* \wedge {\bm l}_*
                  + \omega_2 \,{\bm m}_* \wedge \bar{{\bm m}}_* \,, \label{KY_CCKY_2-form}
\ee
where $\omega_1$ is a real and $\omega_2$ is a pure imaginary function
that satisfy the conditions (\ref{D_KY1})--(\ref{D_KY3}).
Similarly, CCKY 2-forms on type D vacuum solutions must take the same form as (\ref{KY_CCKY_2-form}),
where $\omega_1$ and $\omega_2$ satisfy the conditions
\ba
 (\rho-\bar{\rho})\,\omega_1-(\rho+\bar{\rho})\,\omega_2=0 \,, \label{D_CCKY1}\\
 (\mu-\bar{\mu})\omega_1-(\mu+\bar{\mu})\omega_2=0 \,, \label{D_CCKY2}\\
 (\tau-\bar{\pi})\omega_1-(\tau+\bar{\pi})\omega_2=0 \,. \label{D_CCKY3}
\ea
Hence, covariantly constant 2-forms on type D vacuum solutions take the form (\ref{KY_CCKY_2-form})
that satisfy the conditions (\ref{D_KY1})--(\ref{D_KY3}) and (\ref{D_CCKY1})--(\ref{D_CCKY3}).
The conditions only have the trivial solution $\omega_1=\omega_2=0$.
Therefore, we obtain the following proposition.

\begin{prop} \label{prop4}
All type D vacuum solutions do not admit any covariantly constant forms of rank one, two and three.
\end{prop}

\section{Examples}
It would be interesting to compute the Killing curvatures (\ref{KCurv_KY1})-(\ref{KillingCurvatureKY})
and solve the curvature conditions (\ref{CC_KY_1}), (\ref{CC_KY_2}), (\ref{CC_KY_3}) and (\ref{CC_KY_4})
for various physical metrics.
In this section, we thoroughly investigate KY symmetry of
some type D solutions, a cosmological metric and some gravitational instantons in four dimensions.
We also investigate black hole, ring and string metrics in five dimensions.

The procedure is as follows:
After we compute the Killing curvatures (\ref{KCurv_KY1}) and (\ref{KillingCurvatureKY}),
we first solve the curvature conditions (\ref{CC_KY_1}) and (\ref{CC_KY_2}).
Next, with the solutions of (\ref{CC_KY_1}) and (\ref{CC_KY_2}),
we solve the curvature conditions (\ref{CC_KY_3}) and (\ref{CC_KY_4}).
Finally, using the solutions of (\ref{CC_KY_1}), (\ref{CC_KY_2}), (\ref{CC_KY_3}) and (\ref{CC_KY_4}) as an ansatz,
we attempt to integrate the Killing equations.
Thus, we obtain the precise number of Killing vector fields and KY tensors
for every metric in this section.
The obtained results are listed in Table 1, 2 and 3\footnote{
We have computed the Killing curvatures and solved the curvature conditions
with a package of a computational software ``{\it Mathematica}".
The Mathematica package, we developed by ourselves,
is available on the webpage at {\tt http://www.research.kobe-u.ac.jp/fsci-pacos/KY\_upperbound/}.
}.

\subsection{Type D solutions in four dimensions}
The Plebanski-Demianski solution \cite{Plebanski:1976} is a type D solution
of the Einstein-Maxwell equation with the cosmological constant $\lambda$.
The metric is written in the form \cite{Griffiths:2006},
\be \eqalign{
 ds^2 =& \frac{1}{(1-b pr)^2}\Bigg\{ -\frac{{\cal Q}}{r^2+p^2}(d\tau-p^2d\sigma)^2 \\
       & +\frac{r^2+p^2}{{\cal Q}}dr^2+\frac{r^2+p^2}{{\cal P}}dp^2
          +\frac{{\cal P}}{r^2+p^2}(d\tau+r^2d\sigma)^2 \Bigg\} \,, } \label{PD_metric}
\ee
where
\ba
 {\cal Q} = -(a^2b^2+\lambda/3)r^4-2b nr^3+ r^2-2mr+a^2+e^2+g^2 \,, \\
 {\cal P} = -[(a^2+e^2+g^2)b^2+\lambda/3]p^4+2b mp^3-p^2+2np +a^2 \,.
\ea
The solution contains seven non-vanishing parameters $m$, $n$, $e$, $g$, $a$, $b$ and $\lambda$.
Clearly, the metric admits two Killing vector fields $\partial/\partial\tau$ and $\partial/\partial\sigma$.
No rank-2 and rank-3 KY tensor has been known.
Our calculation shows that the metric admits exactly two Killing vector fields and
no rank-2 and rank-3 KY tensor.

Under a limit of spacetime parameters,
the space of KY solutions might increase in dimension but never reduce,
as was stated in \cite{Geroch:1969}.
In the limit $e=g=\lambda=0$, the Plebanski-Demianski metric becomes
case III solution of Kinnersley's classification,
which has same dimension of the space of KY solutions as the Plebanski-Demianski metric.
This implies that the vanishing of $e$, $g$ and $\lambda$
is not irrespetive of an enhancement of KY symmetry.
However, the vanishing of other parameters does causes an enhancement of KY symmetry.
If the acceleration parameter $b$ is vanishing,
the Plebanski-Demianski metric (\ref{PD_metric}) reduces to
the Carter metric \cite{Carter:1968,Plebanski:1975},
\be \eqalign{
 ds^2 =& \frac{r^2+p^2}{{\cal Q}}dr^2+\frac{r^2+p^2}{{\cal P}}dp^2 \nonumber\\
       & -\frac{{\cal Q}}{r^2+p^2}(d\tau-p^2d\sigma)^2
          +\frac{{\cal P}}{r^2+p^2}(d\tau+r^2d\sigma)^2 \,, } \label{Carter_metric}
\ee
where
\ba
 {\cal Q} = -\lambda/3r^4+ r^2 - 2mr + a^2+e^2+g^2 \,, \\
 {\cal P} = -\lambda/3p^4- p^2 + 2np + a^2 \,.
\ea
The metric admits two Killing vector fields $\partial/\partial\tau$ and $\partial/\partial\sigma$.
It is known \cite{Dietz:1982} that any metric admitting a nondegenerate rank-2 KY tensor 
is written in the form (\ref{Carter_metric}).
Hence, Carter metric admits at least one rank-2 KY tensor.
Our calculation shows that for arbitrary functions ${\cal Q}(r)$ and ${\cal P}(p)$,
the metric admits exactly two Killing vector fields and exactly one rank-2 and no rank-3 KY tensor.
Due to the vanishing of the acceleration parameter $b$,
the rank-2 KY tensor has appeared.

Kerr metric (\ref{Kerr_metric}) recovers when we take $n=e=g=\lambda=0$
in the metric (\ref{Carter_metric}).
Furthermore, we take the static limit of the Kerr metric.
Then, the metric becomes the Schwarzschild metric,
\be
 ds^2 = - f(r) dt^2 + \frac{dr^2}{f(r)} + r^2 d\theta^2 + r^2\sin^2\theta d\phi^2 \,,
\ee
where
\be
 f(r) = 1 - \frac{2m}{r} \,.
\ee
Since the spacetime is static and spherically symmetric, the metric admits four Killing vector fields.
It is also known that the metric admits a (degenerate) rank-2 KY tensor.
Our calculation proves that the metric admits exactly four Killing vector fields,
exactly one rank-2 and no rank-3 KY tensors.

The Wahlquist metric \cite{Wahlquist:1968,Kramer:1985,Senovilla:1987,Wahlquist:1992}
is a type D solution of the Einstein equation for perfect fluids
with the equation of state $\rho+3p=\textrm{const}$.
The metric is given by
\be \eqalign{
 ds^2 =& \frac{r^2+p^2}{{\cal Q}(1+\beta r^2)}dr^2 +\frac{r^2+p^2}{{\cal P}(1+\beta p^2)}dp^2 \\
       & -\frac{{\cal Q}}{r^2+p^2}(d\tau-p^2d\sigma)^2
          +\frac{{\cal P}}{r^2+p^2}(d\tau+r^2d\sigma)^2 \,, } \label{Wahlquist_metric}
\ee
where
\ba
 {\cal Q} =& a^2 -2m r\sqrt{1-\beta^2r^2} + r^2 \nonumber\\
           & + \frac{\mu_0}{\beta^2}\left[r^2-\frac{r \textrm{Arcsin}(\beta r)\sqrt{1-\beta^2r^2}}{\beta}\right] \,, \\
 {\cal P} =& a^2 +2n p\sqrt{1+\beta^2p^2} - p^2 \nonumber\\
           & - \frac{\mu_0}{\beta^2}\left[p^2-\frac{p \textrm{Arcsinh}(\beta p) \sqrt{1+\beta^2p^2}}{\beta}\right] \,.
\ea
The metric contains five constants $m$, $n$, $a$, $\beta$ and $\mu_0$.
One can immediately find that the metric admits two Killing vector fields $\partial/\partial\tau$ and $\partial/\partial\sigma$.
If one takes the limit $\beta\to 0$, the metric becomes the Kerr-NUT-(A)dS metric \cite{Carter:1968}.
For the Wahlquist metric, our calculation shows that it admits
exactly two Killing vector fields, no rank-2 and rank-3 KY tensor.
Note that the Wahlquist metric
admits a rank-2 generalised Killing-Yano tensor with torsion \cite{Hinoue:2014}.

\begin{table}[t]
\begin{center}
\begin{tabular}{lccc} \hline
                  &\multicolumn{3}{c}{$\dim KY^p(M)$} \\
4D metrics        &$p=1$     &$p=2$     &$p=3$        \\ \hline
Maximally symmetric space
                  &10        &10        &5            \\[0.1cm]
Plebanski-Demianski
                  &2         &0         &0            \\
Carter / Kerr  	  &2         &1         &0            \\
Schwarzschild     &4         &1         &0            \\
Wahlquist         &2         &0         &0            \\[0.1cm]
Friedmann-Lema\^itre-Robertson-Walker
                  &6         &4         &1            \\ \hline
\end{tabular}
\caption{The numbers of rank-p KY tensors on some physical spacetimes in four dimensions.}
\end{center}
\end{table}

\subsection{Cosmological model}
The Friedmann-Lema\^itre-Robertson-Walker metric is widely used for a cosmological model.
The metric is given by
\be
 ds^2 = -dt^2 + a(t)^2 \left(\frac{dr^2}{1-Kr^2}+r^2d\theta^2+r^2\sin^2\theta d\phi^2\right) \,, \label{FLRW_metric}
\ee
where $a(t)$ is an arbitrary nonzero function of the time $t$.
Since the spatial part is a three-dimensional maximally symmetric space
with the constant curvature $K$, the metric admits at least six Killing vector fields.
For particular choices of $a(t)$ and $K$,
the FLRW metric describes maximally symmetric,
where the metric admits the maximum number of Killing vector fields.
Our calculation shows that such enhancement of isometry happens
only when the spacetime becomes maximally symmetric.
Calculating the curvature condition, we find that 
the enhancement of isometry happens only when the condition
\be
 a \ddot{a} - \dot{a}^2 - K = 0
\ee
is satisfied.
This shows that the FLRW spacetime becomes maximally symmetric.

A rank-2 KY tensor ${\bm k}$
on $(M^4,{\bm g}_4=-dt^2+a(t)^2\tilde{{\bm g}}_3)$
is given by ${\bm k}=a(t)^3\tilde{{\bm k}}$,
where $\tilde{{\bm k}}$ is a rank-2 KY tensor
on $(M^3,\tilde{{\bm g}}_3)$.
For the FLRW metric, $(M^3,\tilde{{\bm g}}_3)$
is a three-dimensional maximally symmetric space with the constant curvature $K$,
which admits four rank-2 KY tensors.
In a similar fashion to Killing vector fields, our calculation confirms that
the FLRW metric admits the only four rank-2 KY tensors with the exception of maximally symmetric case.
It is also known \cite{Batista:2014} that the metric admits a rank-3 KY tensor.
Our calculation shows that it is the only rank-3 KY tensor in the FLRW spacetime
with the exception of maximally symmetric case.

\subsection{Gravitational instantons}
We discuss Euclidean metrics with the self-dual Weyl curvature in four dimensions,
which are obtained from the Plebanski-Demianski family (\ref{PD_metric}).
We make a shift from $a^2$ to $a^2-n^2$ and then take $n=im$ and $e=ig$.
Moreover, we perform the Wick rotation $r\to ir$ with $m\to im$ and $b\to ib$.
Then, the metric becomes
\be \eqalign{
 ds^2 =& \frac{1}{(1+b pr)^2}\Bigg\{ \frac{{\cal Q}}{r^2-p^2}(d\tau-p^2d\sigma)^2 \\
       & +\frac{r^2-p^2}{{\cal Q}}dr^2+\frac{p^2-r^2}{{\cal P}}dp^2
          +\frac{{\cal P}}{p^2-r^2}(d\tau-r^2d\sigma)^2 \Bigg\} \,, } \label{Euclidean_PD_metric}
\ee
where
\be \eqalign{
 {\cal Q} = [(a^2-m^2)b^2-\lambda/3]r^4+2b mr^3 -r^2+2mr+a^2-m^2 \,, \\
 {\cal P} = [(a^2-m^2)b^2-\lambda/3]p^4-2b mp^3 -p^2-2mp+a^2-m^2 \,. }
\ee
The metric is an Einstein metric, that is, a vacuum solution
to the Einstein equation with cosmological constant $\lambda$.
The Weyl curvature is self-dual.
The BPS conditions were discussed in \cite{Klemm:2013}.
The metric admits two Killing vector fields.
Our calculation finds a rank-2 KY tensor
\be \eqalign{
 {\bm \omega} =& \frac{1}{(r+p)(1+bpr)^3}\Bigg\{\Big(b f(r,p)+\lambda p (r+p)\Big) ~dr\wedge (d\tau-p^2d\sigma) \\
               & + 3b{\cal P} ~dr\wedge (d\tau-r^2d\sigma) + 3b{\cal Q} ~dp\wedge (d\tau-p^2d\sigma) \\
               & + \Big(b f(r,p)+\lambda r (r+p)\Big) ~dp\wedge (d\tau-r^2d\sigma) \Bigg\} \,,}
\ee
where
\be \eqalign{
 f(r,p) =& 3(m^2-a^2)\Big\{(1+bpr)^2+b(r^2+p^2)\Big\} \\
         & -3m(1-bpr)(r-p)+\lambda p^2r^2-3pr \,.
 }
\ee
This rank-2 KY tensor is, as far as we know, new.

When we take $b=0$, the metric (\ref{Euclidean_PD_metric}) becomes
self-dual Kerr-bolt metric \cite{Gibbons:1980} with cosmological constant \cite{Santillan:2007}.
Furthermore, when we take $a=0$, the metric becomes self-dual Taub-NUT metric with cosmological constant \cite{Cvetic:2003}.
Appropriately performing a coordinate transformation, the metric is written as
\be \eqalign{
 ds^2 =& \frac{dr^2}{F(r)} + (r^2-m^2)(d\theta^2+\sin^2\theta d\phi^2) \\
       & + 4m^2F(r)(d\psi + \cos\theta d\phi)^2 \,, }
\ee
where
\ba 
 F(r) = \frac{\lambda}{3}\left(\frac{r+m}{r-m}\right)(r_+-r)(r-r_-) \,, \\
 r_\pm = m\pm\sqrt{4m^2+\frac{3}{\lambda}} \,.
\ea
In particular, for $\lambda=0$, the metric reduces to the self-dual Taub-NUT metric \cite{Hawking:1977}.
The self-dual Taub-NUT metric admits four Killing vector fields.
It also admits three covariantly constant 2-forms, which are the hyper-K\"ahler forms.
In addition, there is a fourth rank-2 KY tensor that is not covariantly constant \cite{vanHolten:1995}.
Our calculation proves that the metric admits the only four Killing vector fields,
the only four rank-2 and no rank-3 KY tensor.

If a self-dual metric is Ricci flat, it becomes hyper-K\"ahler in four dimensions,
which admits three covariantly constant hyper-K\"ahler forms.
This explains an increasing of the number of rank-2 KY tensors
when the cosmological constant vanishes.
Therefore, it is nontrivial \cite{Gibbons:1988,vanHolten:1995}
that the Kerr-bolt and the Taub-NUT metrics with $\lambda=0$
admit one additional rank-2 KY tensor, respectively.

\begin{table}[t]
\begin{center}
\begin{tabular}{lccc} \hline
                    &\multicolumn{3}{c}{$\dim KY^p(M)$} \\
4D Euclidean self-dual metrics
                    &$p=1$     &$p=2$     &$p=3$        \\ \hline
                    &          &          &             \\[-0.2cm]
\underline{$\lambda\neq 0$}
                    &          &          &             \\
Plebanski-Demianski
                    &2         &1         &0            \\
Kerr-bolt           &2         &1         &0            \\
Taub-NUT            &4         &1         &0            \\[0.2cm]
\underline{$\lambda= 0$}
                    &          &          &             \\
Plebanski-Demianski
                    &2         &3         &0            \\
Kerr-bolt           &2         &4         &0            \\
Taub-NUT            &4         &4         &0            \\ \hline
\end{tabular}
\caption{The numbers of rank-p KY tensors on some gravitational instantons.}
\end{center}
\end{table}

\subsection{Black holes, rings and strings in five dimensions}
To see if the method works well also in higher dimensions,
we investigate black holes, rings and strings in five dimensions:
Myers-Perry, Emparan-Reall and Kerr string metrics.
They all are known as stationary, axially symmetric solutions
of the vacuum Einstein equation, which admit three Killing vector fields.

The Myers-Perry metric \cite{Myers:1986} is a vacuum solution
describing rotating black holes in an asymptotically flat spacetime.
The metric in five dimensions is given by
\be \eqalign{
 ds^2 =& \frac{r^2+p^2}{{\cal Q}}dr^2+\frac{r^2+p^2}{{\cal P}}dp^2 \nonumber\\
       & -\frac{{\cal Q}}{r^2+p^2}(d\psi_0-p^2d\psi_1)^2
          +\frac{{\cal P}}{r^2+p^2}(d\psi_0+r^2d\psi_1)^2 \\
       & +\frac{ab}{rp}(d\psi_0+(r^2-p^2)d\psi_1-r^2p^2d\psi_2)^2 \,, } \label{MP_metric}
\ee
where
\be
 {\cal Q} = -\frac{(r^2+a^2)(r^2+b^2)}{r^2}-2m \,, \quad
 {\cal P} = \frac{(a^2-p^2)(b^2-p^2)}{p^2} \,.
\ee
The metric contains three parameters $a$, $b$ and $m$.
The three Killing vector fields are given by $\partial/\partial\psi_0$, $\partial/\partial\psi_1$
and $\partial/\partial\psi_2$.
The metric also admits a rank-3 KY tensor \cite{Page:2007,Krtous:2007}.
Our calculation shows that the Myers-Perry metric admits
exactly three Killing vector fields, exactly one rank-3 
and no rank-2 and rank-4 KY tensor.

The Emparan-Reall metric \cite{Emparan:2002} is a vacuum solution
describing singly rotating black rings.
The metric in the ring coordinates is given by
\be
\eqalign{
 ds^2 =& -\frac{F(y)}{F(x)}\left(dt-CR\frac{1+y}{F(y)}d\psi\right) \\
       & +\frac{R^2F(x)}{(x-y)^2}\left(\frac{dx^2}{G(x)}-\frac{dy^2}{G(y)}+\frac{G(x)}{F(x)}d\phi^2-\frac{G(y)}{F(y)}d\psi^2\right)
}
\ee
where
\be
\eqalign{
 F(\xi) = 1+\lambda\xi \,, \quad G(\xi) = (1-\xi^2)(1+\nu \xi) \,, \\
 C = \sqrt{\lambda(\lambda-\nu)\frac{1+\lambda}{1-\lambda}} \,. }
\ee
The parameters are $R$, $\nu$ and $\lambda$,
which are corresponding to the radius and the thickness of the ring.
The metric admits three Killing vector fields $\partial/\partial t$,
$\partial/\partial\phi$ and $\partial/\partial\psi$.
No higher rank KY tensors have been discovered for the metric.
Our calculation shows that the Emparan-Reall metric admits exactly three Killing vector fields 
and no rank-2, rank-3 and rank-4 KY tensor.

The Kerr string metric is also a vacuum solution describing rotating black strings.
The metric is given by
\be \eqalign{
 ds^2 =& \frac{r^2+p^2}{{\cal Q}}dr^2+\frac{r^2+p^2}{{\cal P}}dp^2 \nonumber\\
       & -\frac{{\cal Q}}{r^2+p^2}(d\tau-p^2d\sigma)^2
          +\frac{{\cal P}}{r^2+p^2}(d\tau+r^2d\sigma)^2+d\psi^2 \,, } \label{Kerr_string_metric}
\ee
where ${\cal Q}$ and ${\cal P}$ are given by (\ref{Kerr_function}).
The metric admits three Killing vector fields $\partial/\partial\tau$, $\partial/\partial\sigma$ and $\partial/\partial\psi$.
The metric also admits a rank-2 KY tensor \cite{Houri:2013}.
Furthermore, the dual 1-form of the Killing vector field $\partial/\partial\psi$, i.e., $d\psi$,
is clearly closed. Hence, the 1-form $d\psi$ is covariantly constant.
This means that the Hodge dual of $d\psi$ is a rank-4 KY tensor.
Our calculation shows that the Kerr string metric admits exactly three Killing vector fields,
exactly one rank-2, no rank-3 and exactly one rank-4 KY tensor.

\begin{table}[t]
\begin{center}
\begin{tabular}{lcccc} \hline
                  &\multicolumn{4}{c}{$\dim KY^p(M)$}     \\
5D metrics        &$p=1$    &$p=2$    &$p=3$    &$p=4$    \\ \hline
Maximally symmetric space
                  &15       &20       &15       &6        \\[0.1cm]
Myers-Perry       &3        &0        &1        &0        \\
Emparan-Reall     &3        &0        &0        &0        \\
Kerr string       &3        &1        &0        &1        \\ \hline
\end{tabular}
\caption{The numbers of rank-p KY tensors on black hole,
 ring and string spacetimes in five dimensions.}
\end{center}
\end{table}

\section{Discussions and conclusions}
In this paper, we have shown a simple method for exploring KY symmetry of spacetimes,
which computes an upper bound on the number of KY tensors,
including Killing vector fields and CCKY tensors,
for a given metric by using the curvature conditions of Killing equations.
Discussing some curvature conditions on Killing vector fields in Sec.~2 
and solving them for the Kerr spacetime in Sec.~3,
we have overviewed how the method is applied to obtain Killing vector fields.
In Sec.~4, we have calculated the curvature conditions on KY and CCKY tensors
from the geometric view point.
We have introduced a connection ${\cal D}$, called a Killing connection,
on the vector bundle $E^p(M)\equiv \Lambda^pT^*M \oplus \Lambda^{p+1}T^*M$
whose parallel sections are in one-to-one correspondence with rank-$p$ KY tensors.
Calculating the curvature of the connection,
we have obtained (\ref{CC_KY_1})--(\ref{CC_KY_2}).
Differentiating the curvature conditions,
we have obtained further conditions (\ref{CC_KY_3})--(\ref{CC_KY_4}).
Similarly, we have obtained (\ref{CC_CCKY_1})--(\ref{CC_CCKY_4}) for CCKY tensors.
Since the number of linearly independent solutions of the curvature conditions
puts an upper bound on the number of Killing vector fields, KY and CCKY tensors,
we have provided the explicit expression of the curvature conditions
in terms of the Riemann curvature.

The solutions to rank-$p$ Killing-Yano equations can exist
if the holonomy of the Killing connection ${\cal D}$ lies in some subgroup of $GL(N,R)$
where $N$ is the rank of $E^p(M)$,
whose generators are given by the Killing curvature and its differentials,
${\cal D}^k {\cal R}$, for $0\leq k<\infty$ \cite{Ambrose:1953}.
Therefore, in order to determine the precise number of KY tensors,
one has to compute an infinite series of the curvature conditions.
However, we have found it interesting that the curvature conditions 
obtained from the Killing curvature ($k=0$) and its differential of first order $(k=1)$
are strong only to determine the precise number of KY tensors
at least for metrics in four and five dimensions investigated in Sec.~5 and 6.
We have thus obtained Table 1, 2 and 3.

The method has some notable features.
The first point is that the curvature conditions are linear algebraic equations,
which enables us to compute the upper bound on the number of KY tensors for any metric.
Another feature is that the method gives an ansatz for solving Killing equations.
Actually, as was seen in Sec.~3, the Killing equation for the Kerr metric
becomes tractable with such an ansatz.
Furthermore, the method applies well also to degenerate cases,
e.g., to the case of Schwarzschild metric where the rank-2 KY tensor is degenerate,
while the rank-2 CCKY nondegenerate case has been well understood
in \cite{Houri:2007,Houri:2008,Krtous:2008,Houri:2009}.
Similarly, the method is also powerful in covariantly constant cases.
Thus, the method discussed in this paper is practically useful
for finding KY tensors for a given metric.

As a future work, it is of great interest to study the curvature conditions on conformal Killing-Yano (CKY) tensors
\cite{Tachibana:1969, Kashiwada:1968}.
Following \cite{Semmelmann:2002,Gover:2008}, one can introduce a Killing connection
whose parallel sections are one-to-one corresponding to CKY tensors.
In a similar way to this paper, one can calculate the curvature of such a connection \cite{Leitner:2004,Houri:2014b}.
The curvature conditions were partly obtained in \cite{Kashiwada:1968,Semmelmann:2002,Stromenger:2010,Batista:2014b}.
As was discussed in \cite{Leitner:2004,Gover:2008,Batista:2014b},
the curvature conditions on CKY tensors are conformal invariant,
which are entirely written in terms of the Weyl curvature and its covariant derivatives.
Similar to this paper, one could discuss the curvature conditions on CKY tensors for type D vacuum spacetimes,
which would give the counterparts to propositions 5.1 to 5.3.
It would be also interesting to investigate CKY symmetry for various metrics,
e.g, see \cite{Mitsuka:2012} for CKY tensors in the near horizon extreme Kerr (NHEK) geometry.

\section*{Acknowledgments}
The authors are grateful to Claude M. Warnick for his useful comments in the early stage of this work.
The authors would also like to thank Kei Yamamoto for his contribution for developing the Mathematica notebook.
Our acknowledgement extends to Carlos Batista, Tohru Eguchi, Yasushi Homma, Masashi Kimura,
David Kubi\v{z}\'nak, Masato Nozawa and Jiro Soda for their helpful comments.
This work was supported by the JSPS Grant-in-Aid for Scientific Research No.\ 26$\cdot$1237
and the Grant-in-Aid for Scientific Research No.\ 23540317.
While T.H. was working at Rikkyo University (until March, 2014),
this work was supported in part by the Research Centre for Measurement in Advanced Science (RCMAS).

\appendix

\section{Type D vacuum conditions in the Newman-Penrose formalism}
To compute the Killing curvature (\ref{KCurv_KY1}) for type D vacuum solutions,
this appendix collects type D vacuum conditions.
When the Riemann curvature defined in (\ref{CurvatureTensor}) acts on $p$-forms,
it is written in the form
\be \eqalign{
 R({\bm X},{\bm Y}) &= (R({\bm X},{\bm Y}){\bm e}^a)\wedge {\bm X}_a \hook \\
                    &= R({\bm X},{\bm Y},{\bm X}^a,{\bm Y}_b)\,\bm{e}^b\wedge {\bm X}_a\hook \,, } \label{Riemann_Operator_2}
\ee
where $R({\bm X},{\bm Y},{\bm Z},{\bm W})=g(R({\bm X},{\bm Y}){\bm Z},{\bm W})$.
The Weyl curvature is defined by
\be \eqalign{
 C({\bm X},{\bm Y},{\bm Z},{\bm W}) =& R({\bm X},{\bm Y},{\bm Z},{\bm W}) \\
                                     & - K({\bm X},{\bm Z})g({\bm Y},{\bm W}) + K({\bm X},{\bm W})g({\bm Y},{\bm Z}) \\
                                     & + K({\bm Y},{\bm Z})g({\bm X},{\bm W}) + K({\bm Y},{\bm W})g({\bm X},{\bm Z}) \,, }
\ee
where $K({\bm X},{\bm Y})$ is the Schouten tensor defined by
\be
 K({\bm X}, {\bm Y}) = \frac{1}{n-2}\left(\frac{S}{2(n-1)}g({\bm X},{\bm Y})-Ric({\bm X},{\bm Y})\right)
\ee
with the Ricci curvature $Ric({\bm X},{\bm Y}) = R({\bm X}^a,{\bm X},{\bm Y},{\bm X}_a)$
and the scalar curvature $S=Ric({\bm X}^a,{\bm X}_a)$.
In analogy with (\ref{Riemann_Operator_2}), we define
\be
 C({\bm X},{\bm Y}) = C({\bm X},{\bm Y},{\bm X}^a,{\bm Y}_b)\,\bm{e}^b\wedge {\bm X}_a\hook \,.
\ee
In type D spacetimes, using the conditions (\ref{curvNP}), we have
\be \eqalign{
 C({\bm k}, {\bm l}) = A({\bm k},{\bm l},{\bm m}) + \bar{A}({\bm k},{\bm l},{\bm m}) \,, \\
 C({\bm k}, {\bm m}) = \Psi (\bar{{\bm m}}_* \wedge {\bm k} \hook + {\bm l}_* \wedge {\bm m} \hook) \,, \\
 C({\bm l}, {\bm m}) = \bar{\Psi} (\bar{{\bm m}}_* \wedge {\bm l} \hook + {\bm k}_* \wedge {\bm m} \hook) \,, \\
 C({\bm m}, \bar{{\bm m}}) = - A({\bm k},{\bm l},{\bm m})+ \bar{A}({\bm k},{\bm l},{\bm m}) \,, }
 \label{W1}
\ee
where 
\be \eqalign{
 A({\bm k},{\bm l},{\bm m})
= \Psi ( {\bm k}_* \wedge {\bm k} \hook - {\bm l}_* \wedge {\bm l} \hook
        + {\bm m}_* \wedge {\bm m} \hook - \bar{{\bm m}}_* \wedge \bar{{\bm m}} \hook) \,, \\
 \bar{A}({\bm k},{\bm l},{\bm m})
= \bar{\Psi} ( {\bm k}_* \wedge {\bm k} \hook - {\bm l}_* \wedge {\bm l} \hook
        - {\bm m}_* \wedge {\bm m} \hook + \bar{{\bm m}}_* \wedge \bar{{\bm m}} \hook) \,. }
\ee

To calculate the covariant derivatives of the Weyl curvature,
we introduce the spin coefficients (\ref{spincoefficients}).
For the null tetrad, the covariant derivatives are calculated as
\be
 \nabla_{{\bm X}_b} {\bm e}^a=- {\bm e}^a(\nabla_{{\bm X}_b} {\bm X}_c) {\bm e}^c \,.
\ee
Hence, we have
\be \eqalign{ 
 \nabla_{{\bm k}}{\bm k}_*
= - \bar{\pi} \,{\bm m}_* - \pi \,\bar{{\bm m}}_* \,, \quad
 \nabla_{{\bm l}}{\bm k}_*
= - (\gamma+\bar{\gamma}) \,{\bm k}_* \,, \\
 \nabla_{{\bm m}}{\bm k}_*
= - (\bar{\alpha}+\beta) \,{\bm k}_* - \mu \,\bar{{\bm m}}_* \,, \\
 \nabla_{{\bm k}}{\bm l}_*
= 0 \,, \quad
 \nabla_{{\bm l}}{\bm l}_*
= (\gamma+\bar{\gamma}) \,{\bm l}_* + \tau \,{\bm m}_* + \bar{\tau} \,\bar{{\bm m}}_* \,, \\
 \nabla_{{\bm m}}{\bm l}_*
= (\bar{\alpha}+\beta) \,{\bm l}_* + \bar{\rho} \,\bar{{\bm m}}_* \,, \\
 \nabla_{{\bm k}}{\bm m}_*
= - \pi \,{\bm l}_* \,, \quad
 \nabla_{{\bm l}}{\bm m}_*
= \bar{\tau} \,{\bm k}_* - (\gamma-\bar{\gamma}) \,{\bm m}_* \,, \\
 \nabla_{{\bm m}}{\bm m}_*
= \bar{\rho} \,{\bm k}_* - \mu \,{\bm l}_* + (\bar{\alpha}-\beta) \,{\bm m}_* \,, \\
 \nabla_{\bar{{\bm m}}}{\bm m}_*
= - (\alpha-\bar{\beta}) \,{\bm m}_* \,, }
\label{s2-1}
\ee
where we have used the conditions (\ref{Goldberg-Sachs}) and $\epsilon=0$.
From the second Bianchi identities, we have the relations
\be \eqalign{
 {\bm k}(\Psi) = 3 \rho \Psi \,, \quad
 {\bm l}(\Psi) = -3 \mu \Psi \,, \\
 {\bm m}(\Psi) = 3 \tau \Psi \,, \quad
 \bar{{\bm m}}(\Psi) = -3 \pi \Psi \,. } \label{s2-2}
\ee
Thanks to the relations (\ref{s2-1})-(\ref{s2-2}),
the covariant derivatives of the Weyl curvature are calculated as
\be \fl \quad \eqalign{
 (\nabla_{{\bm k}}C)({\bm k},{\bm l})
= 3 \rho A({\bm k},{\bm l},{\bm m}) +3 \bar{\rho}\bar{A}({\bm k},{\bm l},{\bm m})
  -3 \pi C({\bm k},{\bm m}) -3 \bar{\pi}C({\bm k},\bar{{\bm m}}), \\
 (\nabla_{{\bm k}}C)({\bm k}, {\bm m})
= 3 \rho C({\bm k}, {\bm m}), \\
 (\nabla_{{\bm k}}C)({\bm l}, {\bm m})
= 3 \bar{\rho}C({\bm l}, {\bm m}) +3 \bar{\pi} \bar{A}({\bm k},{\bm l},{\bm m}),\\
 (\nabla_{{\bm k}}C)({\bm m}, \bar{{\bm m}})
= -3 \rho A({\bm k},{\bm l},{\bm m}) +3 \bar{\rho} \bar{A}({\bm k},{\bm l},{\bm m})
  +3 \pi C({\bm k}, {\bm m}) -3 \bar{\pi}C({\bm k}, \bar{{\bm m}}), \\
 (\nabla_{{\bm l}}C)({\bm k}, {\bm l})
= -3 \mu A({\bm k},{\bm l},{\bm m}) -3 \bar{\mu}\bar{A}({\bm k},{\bm l},{\bm m})
  -3 \tau C({\bm l},\bar{{\bm m}}) -3 \bar{\tau}C({\bm l},{\bm m}), \\
 (\nabla_{{\bm l}}C)({\bm k}, {\bm m})
= -3 \mu C({\bm k}, {\bm m}) +3\tau A({\bm k},{\bm l},{\bm m}),\\
 (\nabla_{{\bm l}}C)({\bm l}, {\bm m})
= -3 \bar{\mu}C({\bm l}, {\bm m}), \\
 (\nabla_{{\bm l}}C)({\bm m}, \bar{{\bm m}})
= 3 \mu A({\bm k},{\bm l},{\bm m}) -3 \bar{\mu} \bar{A}({\bm k},{\bm l},{\bm m})
 +3 \tau C({\bm l}, \bar{{\bm m}}) -3 \bar{\tau}C({\bm l}, {\bm m}), \\ 
 (\nabla_{{\bm m}}C)({\bm k}, {\bm l})
= 3 \tau A({\bm k},{\bm l},{\bm m}) -3 \bar{\pi}\bar{A}({\bm k},{\bm l},{\bm m})
  -3 \mu C({\bm k}, {\bm m}) -3 \bar{\rho}C({\bm l}, {\bm m}), \\
 (\nabla_{{\bm m}}C)({\bm k}, {\bm m})
= 3 \tau C({\bm k}, {\bm m}), \\
 (\nabla_{{\bm m}}C)({\bm l}, {\bm m})
= -3 \bar{\pi}C({\bm l}, {\bm m}), \\
 (\nabla_{{\bm m}}C)({\bm m}, \bar{{\bm m}})
= -3 \tau A({\bm k},{\bm l},{\bm m})-3 \bar{\pi} \bar{A}({\bm k},{\bm l},{\bm m})
  +3 \mu C({\bm k} ,{\bm m}) -3 \bar{\rho}C({\bm l}, {\bm m}) \,,\\
 (\nabla_{\bar{{\bm m}}}C)({\bm k}, {\bm m})
= -3 \pi C({\bm k}, {\bm m}) +3 \rho A({\bm k},{\bm l},{\bm m}),\\
 (\nabla_{\bar{{\bm m}}}C)({\bm l}, {\bm m})
= 3 \bar{\tau}C({\bm l}, {\bm m}) +3\bar{\mu} \bar{A}({\bm k},{\bm l},{\bm m}). } \label{CD_Weyl}
\ee

We have obtained type D vacuum conditions on the Weyl curvature (\ref{W1}) and its covariant derivatives (\ref{CD_Weyl}).
Since the Riemann curvature is replaced by the Weyl curvature in vacuum spacetimes,
we can compute the Killing curvature (\ref{KCurv_KY1}) on $E^p(M)$ for type D vacuum solutions.
In subsequent sections, we provide the explicit expressions of the Killing curvatures
for $p=1$, $2$ and $3$.
Then, we use the following basis on $\Lambda^p(M)$:\\
(a) 1-forms
\be
 {\bm e}^1 = {\bm k}_* \,, \quad {\bm e}^2 = \textrm{\boldmath $\ell$}_* \,, \quad
 {\bm e}^3 = \textrm{\boldmath $m$}_* \,, \quad {\bm e}^4 = \textrm{\boldmath $\bar{m}$}_* \,. \label{Basis_1}
\ee
(b) 2-forms
\ba
 {\bm e}^1_{(2)} = \textrm{\boldmath $k_*$}\wedge  \textrm{\boldmath $\ell_*$} \,, \quad 
 {\bm e}^2_{(2)} = \textrm{\boldmath $k_*$}\wedge  \textrm{\boldmath $m_*$} \,, \quad
 {\bm e}^3_{(2)} = \textrm{\boldmath $k_*$}\wedge  \textrm{\boldmath $\bar{m}_*$} \,, \nonumber\\
 {\bm e}^4_{(2)} = \textrm{\boldmath $\ell_*$}\wedge  \textrm{\boldmath $m_*$} \,, \quad
 {\bm e}^5_{(2)} = \textrm{\boldmath $\ell_*$}\wedge  \textrm{\boldmath $\bar{m}_*$} \,, \quad 
 {\bm e}^6_{(2)} = \textrm{\boldmath $m_*$}\wedge  \textrm{\boldmath $\bar{m}_*$} \,. \label{Basis_2}
\ea
(c) 3-forms
\ba
 {\bm e}^1_{(3)} = {\bm k}_*\wedge {\bm \ell}_*\wedge {\bm m}_* \,, \quad 
 {\bm e}^2_{(3)} = {\bm k}_*\wedge {\bm \ell}_*\wedge \bar{{\bm m}}_* \,, \nonumber\\
 {\bm e}^3_{(3)} = {\bm k}_*\wedge {\bm m}_*\wedge \bar{{\bm m}}_* \,, \quad 
 {\bm e}^4_{(3)} = {\bm \ell}_*\wedge {\bm m}_*\wedge \bar{{\bm m}}_* \,. \label{Basis_3}
\ea
(d) 4-forms
\be
 \textrm{vol} = {\bm k}_*\wedge {\bm l}_*\wedge {\bm m}_*\wedge \bar{{\bm m}}_* \,. \label{Basis_4}
\ee

\subsection{Killing curvature for $p=1$}
Let us first compute the Killing curvature on $E^1(M)$.
Since we have $N_{11}({\bm X},{\bm Y})=0$ for any vectors ${\bm X}$ and ${\bm Y}$,
the curvature condition is given by
\be
 N_{21}({\bm X},{\bm Y}){\bm \omega} +N_{22}({\bm X},{\bm Y}){\bm \eta} = 0 \,. \label{CC1_0}
\ee
Namely, we only have to compute $N_{21}({\bm X},{\bm Y})$ and $N_{22}({\bm X},{\bm Y})$.
With respect to the basis (\ref{Basis_1}) and (\ref{Basis_2}),
the section $\hat{{\bm \omega}}=({\bm \omega},{\bm \eta})$ of $E^1(M)$ is written as
\be
 {\bm \omega} = \sum_{j=1}^4 \omega_j {\bm e}^j \,, \quad
 {\bm \eta} = \sum_{i=1}^6 \eta_i  {\bm e}^i_{(2)} \,,
\ee
where $\omega_j$ and $\eta_i$ are unknown functions.
For every choice of vector basis ${\bm X}$ and ${\bm Y}$,
$N_{21}({\bm X},{\bm Y})$ and $N_{22}({\bm X},{\bm Y})$ are given as matrices by
\be \fl \quad
 N_{21}(\textrm{\boldmath $k$},\textrm{\boldmath $\ell$})
= 6 \left(
\begin{array}{@{\,}cccc@{\,}}
 \mu \Psi+\bar{\mu}\bar{\Psi} &-\rho \Psi-\bar{\rho} \bar{\Psi} &-\pi \Psi+\bar{\tau}\bar{\Psi} &\tau \Psi-\bar{\pi}\bar{\Psi} \\
 -\tau \Psi & 0 & \rho \Psi  & 0 \\
 -\bar{\tau} \bar{\Psi} & 0 & 0 &\bar{\rho} \bar{\Psi} \\
 0 & -\bar{\pi} \bar{\Psi} & \bar{\mu} \bar{\Psi} & 0 \\
 0 & -\pi \Psi & 0 &\mu \Psi \\
 -\mu \Psi+\bar{\mu}\bar{\Psi} &\rho \Psi-\bar{\rho} \bar{\Psi} &\pi \Psi+\bar{\tau}\bar{\Psi} &-\tau \Psi-\bar{\pi}\bar{\Psi}
\end{array}
\right) \,, \label{KC1_1}
\ee
\be
 N_{21}(\textrm{\boldmath $k$},\textrm{\boldmath $m$})
= 6\Psi
\left(
\begin{array}{cccc}
 -\tau &0 &\rho & 0 \\
 0 & 0 & 0 & 0 \\
 0 & 0 & 0 & 0 \\
 0 & 0 & 0 & 0 \\
 -\mu  &\rho  &\pi  & -\tau  \\
 \tau  &0 &- \rho &0 
\end{array}
\right) \,,\label{KC1_2}
\ee
\be
 N_{21}(\textrm{\boldmath $k$},\textrm{\boldmath $\bar{m}$})
= 6\bar{\Psi}
\left(
\begin{array}{cccc}
 -\bar{\tau}  &0 &0 &\bar{\rho} \\
 0 & 0 & 0 & 0 \\
 0 & 0 & 0 & 0 \\
 -\bar{\mu} &\bar{\rho} & -\bar{\tau} & \bar{\pi} \\
 0 & 0 & 0 & 0 \\
 -\bar{\tau} & 0 & 0 & \bar{\rho}
\end{array}
\right) \,, \label{KC1_3}
\ee
\be
 N_{21}(\textrm{\boldmath $\ell$},\textrm{\boldmath $m$})
= 6\bar{\Psi} \left(
\begin{array}{@{\,}cccc@{\,}}
 0 &-\bar{\pi} &\bar{\mu} &0 \\
 0 &0 &0 &0 \\
 -\bar{\mu} &\bar{\rho} & - \bar{\tau} & \bar{\pi} \\
 0 & 0 & 0 & 0 \\
 0 & 0 & 0 & 0 \\
 0 &-\bar{\pi} & \bar{\mu} & 0
\end{array}
\right) \,, \label{KC1_4}
\ee 
\be
 N_{21}(\textrm{\boldmath $\ell$},\textrm{\boldmath $\bar{m}$})
= 6\Psi \left(
\begin{array}{@{\,}cccc@{\,}}
 0 &-\pi &0 &\mu \\
 -\mu &\rho &\pi & -\tau \\
 0 & 0 & 0 & 0 \\
 0 & 0 & 0 & 0 \\
 0 & 0 & 0 & 0 \\
 0 & \pi &0 &- \mu
\end{array}
\right) \,, \label{KC1_5}
\ee
\be \fl \quad
 N_{21}(\textrm{\boldmath $m$},\textrm{\boldmath $\bar{m}$})
= 6
\left(
\begin{array}{@{\,}cccc@{\,}}
 -\mu\Psi+\bar{\mu}\bar{\Psi} & \rho\Psi-\bar{\rho}\bar{\Psi} &\pi\Psi+\bar{\tau}\bar{\Psi} &-\tau\Psi-\bar{\pi}\bar{\Psi} \\
 \tau\Psi & 0 &- \rho\Psi  & 0 \\
 -\bar{\tau}\bar{\Psi} & 0 & 0 & \bar{\rho}\bar{\Psi} \\
 0 & -\bar{\pi}\bar{\Psi} & \bar{\mu}\bar{\Psi} & 0 \\
 0 & \pi\Psi & 0 &-\mu\Psi \\
 \mu\Psi+\bar{\mu}\bar{\Psi} &-\rho\Psi-\bar{\rho}\bar{\Psi} &-\pi\Psi+\bar{\tau}\bar{\Psi} &\tau\Psi-\bar{\pi}\bar{\Psi}
\end{array}
\right) \,, \label{KC1_6}
\ee
\be
 N_{22}(\textrm{\boldmath $k$},\textrm{\boldmath $\ell$})
= 3 \left(
\begin{array}{cccccc}
0 &0     &0           &0            &0      &0 \\
0 &\Psi &0           &0            &0      &0 \\
0 &0     &\bar{\Psi} &0            &0      &0 \\
0 &0     &0           &-\bar{\Psi} &0      &0 \\
0 &0     &0           &0            &-\Psi &0 \\
0 &0     &0           &0            &0      &0
\end{array}
\right) \,, \label{KC1_7}
\ee
\be
 N_{22}(\textrm{\boldmath $k$},\textrm{\boldmath $m$})
= 3 \left(
\begin{array}{cccccc} 
0 &\Psi  &0 &0 &0 &0 \\
0 &0      &0 &0 &0 &0 \\
0 &0      &0 &0 &0 &0 \\
0 &0      &0 &0 &0 &0 \\
0 &0      &0 &0 &0 &0 \\
0 &-\Psi &0 &0 &0 &0
\end{array}
\right) \,, \label{KC1_8}
\ee 
\be
 N_{22}(\textrm{\boldmath $k$},\textrm{\boldmath $\bar{m}$})
= 3\left(
\begin{array}{cccccc}
0 &0 &\bar{\Psi} &0 &0 & 0 \\
0 &0 &0 &0 &0 &0 \\
0 &0 &0 &0 &0 &0 \\
0 &0 &0 &0 &0 &0 \\
0 &0 &0 &0 &0 &0 \\
0 &0 &\bar{\Psi} & 0 &0 & 0
\end{array}
\right) \,, \label{KC1_9}
\ee
\be
 N_{22}(\textrm{\boldmath $\ell$},\textrm{\boldmath $m$})
= 3\left(
\begin{array}{cccccc}
0 &0 &0 &-\bar{\Psi} &0 &0 \\
0 &0 &0 &0             &0 &0 \\
0 &0 &0 &0             &0 &0 \\
0 &0 &0 &0             &0 &0 \\
0 &0 &0 &0             &0 &0 \\
0 &0 &0 &-\bar{\Psi} &0 &0
\end{array}
\right) \,, \label{KC1_10}
\ee
\be
 N_{22}(\textrm{\boldmath $\ell$},\textrm{\boldmath $\bar{m}$})
= 3\left(
\begin{array}{cccccc}
0 &0 &0 &0 &-\Psi &0 \\
0 &0 &0 &0 &0      &0 \\
0 &0 &0 &0 &0      &0 \\
0 &0 &0 &0 &0      &0 \\
0 &0 &0 &0 &0      &0 \\
0 &0 &0 &0 & \Psi &0
\end{array}
\right) \,, \label{KC1_11}
\ee
\be
 N_{22}(\textrm{\boldmath $m$},\textrm{\boldmath $\bar{m}$})
= 3\left(
\begin{array}{cccccc}
0 &0      &0           &0            &0     &0 \\
0 &-\Psi &0           &0            &0     &0 \\
0 &0      &\bar{\Psi} &0            &0     &0 \\
0 &0      &0           &-\bar{\Psi} &0     &0 \\
0 &0      &0           &0            &\Psi &0 \\
0 &0      &0           &0            &0     &0
\end{array}
\right) \,. \label{KC1_12}
\ee

\subsection{Killing curvature for $p=2$}
Next, we compute the Killing curvature on $E^2(M)$.
The curvature condition is given by
\ba
 \left(
\begin{array}{cc}
N_{11}({\bm X},{\bm Y}) &0  \\
N_{12}({\bm X},{\bm Y}) &N_{22}({\bm X},{\bm Y}) 
\end{array}
\right)
\left(
\begin{array}{c}
 {\bm \omega} \\
 {\bm \eta}
\end{array}
\right) = 0 \,, \label{CC2_0}
\ea
With respect to the basis (\ref{Basis_2}) and (\ref{Basis_3}),
the section $\hat{{\bm \omega}}=({\bm \omega},{\bm \eta})$ of $E^1(M)$ is written as
\be
 {\bm \omega} = \sum_{j=1}^6 \omega_j {\bm e}^j_{(2)} \,, \quad
 {\bm \eta}   = \sum_{i=1}^4 \eta_i  {\bm e}^i_{(3)} \,,
\ee
where $\omega_j$ and $\eta_i$ are unknown functions.
For every choice of vector basis ${\bm X}$ and ${\bm Y}$,
the matrices of $N_{11}({\bm X},{\bm Y})$, $N_{21}({\bm X},{\bm Y})$ and $N_{22}({\bm X},{\bm Y})$
with respect to the basis (\ref{Basis_2}) and (\ref{Basis_3}) are given by
\be
 N_{11}(\textrm{\boldmath $k$},\textrm{\boldmath $\ell$})
= \frac{3}{2} \left(
\begin{array}{cccccc}
0 &0 &0 &0 &0 &0 \\
0 &\Psi &0 &0 &0 &0 \\
0 &0 &\bar{\Psi} &0 &0 &0 \\
0 &0 &0 &-\bar{\Psi} &0 &0 \\
0 &0 &0 &0 &-\Psi &0 \\
0 &0 &0 &0 &0 &0
\end{array}
\right) \,, \label{KC2_1}
\ee
\be
 N_{11}(\textrm{\boldmath $k$},\textrm{\boldmath $m$})
= \frac{3}{2} \left(
\begin{array}{cccccc}
0 &\Psi &0 &0 &0 &0 \\
0 &0 &0 &0 &0 &0 \\
0 &0 &0 &0 &0 &0 \\
0 &0 &0 &0 &0 &0 \\
0 &0 &0 &0 &0 &0 \\
0 &-\Psi &0 &0 &0 &0
\end{array}
\right) \,, \label{KC2_2}
\ee
\be
 N_{11}(\textrm{\boldmath $k$},\textrm{\boldmath $\bar{m}$})
= \frac{3}{2}\left(
\begin{array}{cccccc}
0 &0 &\bar{\Psi} &0 &0 &0 \\
0 &0 &0 &0 &0 &0 \\
0 &0 &0 &0 &0 &0 \\
0 &0 &0 &0 &0 &0 \\
0 &0 &0 &0 &0 &0 \\
0 &0 &\bar{\Psi} &0 &0 &0
\end{array}
\right) \,, \label{KC2_3}
\ee
\be
 N_{11}(\textrm{\boldmath $\ell$},\textrm{\boldmath $m$})
= \frac{3}{2}\left(
\begin{array}{cccccc}
0 &0 &0 &-\bar{\Psi} &0 &0 \\
0 &0 &0 &0 &0 &0 \\
0 &0 &0 &0 &0 &0 \\
0 &0 &0 &0 &0 &0 \\
0 &0 &0 &0 &0 &0 \\
0 &0 &0 &-\bar{\Psi} &0 &0
\end{array}
\right) \,, \label{KC2_4}
\ee
\be
 N_{11}(\textrm{\boldmath $\ell$},\textrm{\boldmath $\bar{m}$})
= \frac{3}{2} \left(
\begin{array}{cccccc}
0 &0 &0 &0 &-\Psi &0 \\
0 &0 &0 &0 &0 &0 \\
0 &0 &0 &0 &0 &0\\
0 &0 &0 &0 &0 &0 \\
0 &0 &0 &0 &0 &0 \\
0 &0 &0 &0 &\Psi &0
\end{array}
\right) \,, \label{KC2_5}
\ee
\be
 N_{11}(\textrm{\boldmath $m$},\textrm{\boldmath $\bar{m}$})
= \frac{3}{2} \left(
\begin{array}{@{\,}cccccc@{\,}}
0 &0 &0 &0 &0 &0 \\
0 &-\Psi &0 &0 &0 &0 \\
0 &0 &\bar{\Psi} &0 &0 &0 \\
0 &0 &0 &-\bar{\Psi} &0 &0 \\
0 &0 &0 &0 &\Psi &0 \\
0 &0 &0 &0 &0 &0
\end{array}
\right) \,, \label{KC2_6}
\ee
\be \fl \quad
 N_{21}(\textrm{\boldmath $k$},\textrm{\boldmath $\ell$})
= \frac{9}{2}\left(
\begin{array}{cccccc}
 \tau \Psi+\bar{\pi}\bar{\Psi} &\mu \Psi &0 &-\bar{\rho}\bar{\Psi} &0 &-\tau \Psi+\bar{\pi}\bar{\Psi} \\
 \pi\Psi+\bar{\tau}\bar{\Psi} &0 &\bar{\mu}\bar{\Psi} &0 &-\rho \Psi  &-\pi \Psi+\bar{\tau}\bar{\Psi} \\
 -\rho \Psi+\bar{\rho}\bar{\Psi} &-\pi \Psi &\bar{\pi}\bar{\Psi} &0 &0 &\rho \Psi+\bar{\rho}\bar{\Psi} \\
-\mu \Psi+\bar{\mu} \bar{\Psi} & 0 & 0 &-\bar{\tau} \bar{\Psi} & \tau \Psi & \mu \Psi+\bar{\mu} \bar{\Psi}
\end{array}
\right) \,, \label{KC2_7}
\ee 
\be
 N_{21}(\textrm{\boldmath $k$},\textrm{\boldmath $m$})
= \frac{9}{2}\Psi \left(
\begin{array}{cccccc}
0 &-\tau &0 &0 &0 &0 \\
-\rho &-\pi &0 &0 &0 &\rho \\
0 &\rho &0 &0 &0 &0 \\
\tau &\mu &0 &0 &0 &-\tau
\end{array}
\right) \,, \label{KC2_8}
\ee
\be
 N_{21}(\textrm{\boldmath $k$},\textrm{\boldmath $\bar{m}$})
= \frac{9}{2}\bar{\Psi} \left(
\begin{array}{cccccc}
-\bar{\rho} &0 &-\bar{\pi} &0 &0 &-\bar{\rho} \\
0 &0 &-\bar{\tau} &0 &0 &0 \\
0 &0 &-\bar{\rho} &0 &0 &0 \\
-\bar{\tau} &0 &-\bar{\mu} &0 &0 &-\bar{\tau}
\end{array}
\right) \,, \label{KC2_9}
\ee
\be
 N_{21}(\textrm{\boldmath $\ell$},\textrm{\boldmath $m$})
= \frac{9}{2}\bar{\Psi} \left(
\begin{array}{cccccc}
0 &0 &0 &-\bar{\pi} &0 &0 \\
\bar{\mu} &0 &0 &-\bar{\tau} &0 &\bar{\mu} \\
\bar{\pi} &0 &0 &-\bar{\rho} &0 &\bar{\pi} \\
0 &0 &0 &-\bar{\mu} &0 &0 
\end{array}
\right) \,, \label{KC2_10}
\ee
\be
 N_{21}(\textrm{\boldmath $\ell$},\textrm{\boldmath $\bar{m}$})
= \frac{9}{2}\Psi\left(
\begin{array}{cccccc}
\mu &0 &0 &0 &-\tau &-\mu \\
0 &0 &0 &0 &-\pi &0 \\
-\pi &0 &0 &0 &\rho &\pi \\
0 &0 &0 &0 &\mu &0
\end{array}
\right) \,, \label{KC2_11}
\ee
\be \fl \quad
 N_{21}(\textrm{\boldmath $m$},\textrm{\boldmath $\bar{m}$})
=\frac{9}{2} \left(
\begin{array}{@{\,}cccccc@{\,}}
-\tau \Psi+\bar{\pi}\bar{\Psi} &-\mu \Psi &0 &-\bar{\rho}\bar{\Psi} &0 &\tau \Psi+\bar{\pi}\bar{\Psi} \\
-\pi \Psi+\bar{\tau}\bar{\Psi} &0 &\bar{\mu}\bar{\Psi} &0 &\rho &\pi \Psi+\bar{\tau}\bar{\Psi} \\
\rho \Psi+\bar{\rho}\bar{\Psi} &\pi \Psi &\bar{\pi}\bar{\Psi} &0 &0 &-\rho \Psi+\bar{\rho}\bar{\Psi} \\
\mu \Psi+\bar{\mu} \bar{\Psi} &0 &0 &-\bar{\tau} \bar{\Psi} &-\tau \Psi &-\mu \Psi+\bar{\mu} \bar{\Psi}
\end{array}
\right) \,, \label{KC2_12}
\ee
\be
 N_{22}(\textrm{\boldmath $k$},\textrm{\boldmath $\ell$})
= \frac{3}{2} \left(
\begin{array}{cccc}
\Psi-\bar{\Psi} &0 &0 &0 \\
0 &-\Psi+\bar{\Psi} &0 &0 \\
0 &0 &\Psi+\bar{\Psi} & 0\\
0 &0 &0 &-\Psi-\bar{\Psi}
\end{array}
\right) \,, \label{KC2_13}
\ee
\be
 N_{22}(\textrm{\boldmath $k$},\textrm{\boldmath $m$})
= \frac{3}{2} \left(
\begin{array}{cccc}
0 &0 &0 &0 \\
0 &0 &\Psi &0 \\
0 &0 &0 &0 \\
\Psi &0 &0 &0 
\end{array}
\right) \,, \label{KC2_14}
\ee
\be
 N_{22}(\textrm{\boldmath $k$},\textrm{\boldmath $\bar{m}$})
= \frac{3}{2} \left(
\begin{array}{cccc}
0 &0 &-\bar{\Psi} &0 \\
0 &0 &0 &0 \\
0 &0 &0 &0 \\
0 & -\bar{\Psi} &0 &0 
\end{array}
\right) \,, \label{KC2_15}
\ee
\be
 N_{22}(\textrm{\boldmath $\ell$},\textrm{\boldmath $m$})
= \frac{3}{2} \left(
\begin{array}{cccc}
0 &0 &0 &0 \\
0 &0 &0 &-\bar{\Psi} \\
-\bar{\Psi} &0 &0 &0 \\
0 &0 &0 &0 
\end{array}
\right) \,, \label{KC2_16}
\ee
\be
 N_{22}(\textrm{\boldmath $\ell$},\textrm{\boldmath $\bar{m}$})
= \frac{3}{2} \left(
\begin{array}{cccc}
0 &0 &0 &\Psi \\
0 &0 &0 &0 \\
0 &\Psi &0 &0 \\
0 &0 &0 &0 
\end{array}
\right) \,, \label{KC2_17}
\ee
\be
 N_{22}(\textrm{\boldmath $m$},\textrm{\boldmath $\bar{m}$})
= \frac{3}{2} \left(
\begin{array}{cccc}
-\Psi-\bar{\Psi} &0 &0 &0 \\
0 &\Psi+\bar{\Psi} &0 &0 \\
0 &0 &-\Psi+\bar{\Psi} &0\\
0 &0 &0 &\Psi-\bar{\Psi}
\end{array}
\right) \,. \label{KC2_18}
\ee

\subsection{Killing curvature for $p=3$}
Finally, we compute the Killing curvature on $E^3(M)$.
Now, we have $N_{22}({\bm X},{\bm Y})=0$.
Since we also have $N_{21}({\bm X},{\bm Y})=0$,
the curvature condition is given by
\be
 N_{11}({\bm X},{\bm Y}){\bm \omega} = 0 \,. \label{CC3_0}
\ee
With respect to the basis (\ref{Basis_3}) and (\ref{Basis_4}), 
the section $\hat{{\bm \omega}}=({\bm \omega},{\bm \eta})$ of $E^3(M)$
is written as
\be
 {\bm \omega} = \sum_{j=1}^4 \omega_j {\bm e}^j_{(3)} \,, \quad
 {\bm \eta}   = \eta_1 \textrm{vol} \,.
\ee
where $\omega_j$ and $\eta_1$ are unknown functions.
For every choice of basis ${\bm X}$ and ${\bm Y}$,
the matrix of $N_{11}({\bm X},{\bm Y})$ is given by
\be
 N_{11}(\textrm{\boldmath $k$},\textrm{\boldmath $\ell$})
= \left(
\begin{array}{cccc}
\Psi-\bar{\Psi} &0 &0 &0 \\
0 &-\Psi+\bar{\Psi} &0 &0 \\
0 &0 &\Psi+\bar{\Psi} &0 \\
0 &0 &0 &-\Psi-\bar{\Psi}
\end{array}
\right) \,, \label{KC3_1}
\ee
\be
 N_{11}(\textrm{\boldmath $k$},\textrm{\boldmath $m$})
= \left(
\begin{array}{cccc}
0 &0 &0 &0 \\
0 &0 &\Psi &0 \\
0 &0 &0 &0 \\
\Psi &0 &0 &0 
\end{array}
\right) \,, \label{KC3_2}
\ee
\be
 N_{11}(\textrm{\boldmath $k$},\textrm{\boldmath $\bar{m}$})
= \left(
\begin{array}{cccc}
0 &0 &-\bar{\Psi} &0 \\
0 &0 &0 &0 \\
0 &0 &0 &0 \\
0 &-\bar{\Psi} &0 &0 
\end{array}
\right) \,, \label{KC3_3}
\ee
\be
 N_{11}(\textrm{\boldmath $\ell$},\textrm{\boldmath $m$})
= \left(
\begin{array}{cccc}
0 &0 &0 &0 \\
0 &0 &0 &-\bar{\Psi} \\
-\bar{\Psi} &0 &0 &0 \\
0 &0 &0 &0 
\end{array}
\right) \,, \label{KC3_4}
\ee
\be
 N_{11}(\textrm{\boldmath $\ell$},\textrm{\boldmath $\bar{m}$})
= \left(
\begin{array}{@{\,}cccc@{\,}}
0 &0 &0 &\Psi \\
0 &0 &0 &0 \\
0 &\Psi &0 &0 \\
0 &0 &0 &0 
\end{array}
\right) \,, \label{KC3_5}
\ee
\be
 N_{11}(\textrm{\boldmath $m$},\textrm{\boldmath $\bar{m}$})
= \left(
\begin{array}{@{\,}cccc@{\,}}
-\Psi-\bar{\Psi} &0 &0 &0 \\
0 &\Psi+\bar{\Psi} &0 &0 \\
0 &0 &-\Psi+\bar{\Psi} &0 \\
0 &0 &0 &\Psi-\bar{\Psi}
\end{array}
\right) \,. \label{KC3_6}
\ee

\section*{References}

\end{document}